\documentstyle[12pt]{article} 

\def\doit#1#2{\ifcase#1\or#2\fi} 

\expandafter\ifx\csname amsppt.sty\endcsname\endinput
  \expandafter\def\csname amsppt.sty\endcsname{2.2 (2001/08/07)}\fi

\catcode`@=11 
\catcode`@=12 

\let\du=\d                      

\def\a{\alpha} \def\b{\beta}  \def\d{\delta}
\def\e{\epsilon}  \def\g{\gamma}
   \def\k{\kappa}
\def\l{\lambda} \def\m{\mu} \def\n{\nu} \def\o{\omega}
  \def\r{\rho} \def\s{\sigma}
\def\t{\tau}

\def\pmb#1{\setbox0=\hbox{${#1}$}%
   \kern-.025em\copy0\kern-\wd0
   \kern-.035em\copy0\kern-\wd0
   \kern.05em\copy0\kern-\wd0
   \kern-.035em\copy0\kern-\wd0
   \kern-.025em\box0 }

\def\bo{{\raise-.46ex\hbox{\large$\Box$}}} 

\def\pr{\prod}                            

\def\TH{{\raise.2ex\hbox{$\displaystyle \bigodot$}\mskip-4.7mu %
\llap H \;}}
\def\face{{\raise.2ex\hbox{$\displaystyle \bigodot$}\mskip-2.2mu %
\llap {$\ddot
        \smile$}}}                           

\def\sp#1{{}^{#1}}                 

\def\Tilde#1{{\widetilde{#1}}\hskip 0.015in}     
\def\Hat#1{\widehat{#1}}                        
\def\Bar#1{\overline{#1}}                       
\def\leftrightarrowfill{$\mathsurround=0pt \mathord\leftarrow 
 \mkern-6mu
        \cleaders\hbox{$\mkern-2mu \mathord- \mkern-2mu$}\hfill
        \mkern-6mu \mathord\rightarrow$}
\def\dvec#1{\vbox{\ialign{##\crcr
        \leftrightarrowfill\crcr\noalign{\kern-1pt\nointerlineskip}
        $\hfil\displaystyle{#1}\hfil$\crcr}}}           
\def\dt#1{{\buildrel {\hbox{\LARGE .}} \over {#1}}}

\def\frac#1#2{{\textstyle{#1\over\vphantom2\smash{\raise.20ex
        \hbox{$\scriptstyle{#2}$}}}}}   
\def\sfrac#1#2{{\vphantom1\smash{\lower.5ex\hbox{\small$#1$}}\over
        \vphantom1\smash{\raise.4ex\hbox{\small$#2$}}}}
\def\bfrac#1#2{{\vphantom1\smash{\lower.5ex\hbox{$#1$}}\over
        \vphantom1\smash{\raise.3ex\hbox{$#2$}}}}       
\def\afrac#1#2{{\vphantom1\smash{\lower.5ex\hbox{$#1$}}\over#2}} 
\def\on#1#2{\mathop{\null#2}\limits^{#1}}       

\newskip\humongous \humongous=0pt plus 1000pt minus 1000pt
\def\caja{\mathsurround=0pt}

\newif\ifdtup
\def\panorama{\global\dtuptrue \openup2\jot \caja
        \everycr{\noalign{\ifdtup \global\dtupfalse
        \vskip-\lineskiplimit \vskip\normallineskiplimit
        \else \penalty\interdisplaylinepenalty \fi}}}
\def\li#1{\panorama \tabskip=\humongous      
        \halign to\displaywidth{\hfil$\displaystyle{##}$
        \tabskip=0pt&$\displaystyle{{}##}$\hfil
        \tabskip=\humongous&\llap{$##$}\tabskip=0pt
        \crcr#1\crcr}}

\doit0{
\def\ref#1{$\sp{#1)}$}
}

\parskip=\medskipamount          
\def\baselinestretch{1.2}       
\thicklines                         

\def\endtitle{\end{quotation}\newpage}  

\def\sect#1{\bigskip\medskip \goodbreak \noindent{\bf {#1}} %
\nobreak \medskip}
\def\refs{\sect{References} \footnotesize \frenchspacing \parskip=0pt}
\def\Item{\par\hang\textindent}

\def\[{\lfloor{\hskip 0.35pt}\!\!\!\lceil}
\def\]{\rfloor{\hskip 0.35pt}\!\!\!\rceil}

\def\Lag{{\cal L}}
\def\du#1#2{_{#1}{}^{#2}}

\def\calF{{\cal F}}
\def\calH{{\cal H}}

\def\rma{{\rm a}} \def\rmb{{\rm b}} \def\rmc{{\rm c}} 
\def\rmd{{\rm d}} 
\def\rme{{\rm e}} \def\rmf{{\rm f}} \def\rmg{{\rm g}} 
\def\rmh{{\rm h}}  \def\rmi{{\rm i}}

\def\plpl{{+\!\!\!\!\!{\hskip 0.009in}%
{\raise-1.0pt\hbox{$_+$}}  {\hskip 0.0008in}}} 
\def\mimi{{-\!\!\!\!\!{\hskip 0.009in}%
{\raise-1.0pt\hbox{$_-$}}  {\hskip 0.0008in}}}

\def\order#1#2{{\cal O}({#1}^{#2})}

\def\pl#1#2#3{Phys.~Lett.~{\bf {#1}B} (19{#2}) #3}
\def\np#1#2#3{Nucl.~Phys.~{\bf B{#1}} (19{#2}) #3}
\def\prl#1#2#3{Phys.~Rev.~Lett.~{\bf #1} (19{#2}) #3}
\def\pr#1#2#3{Phys.~Rev.~{\bf D{#1}} (19{#2}) #3}

\def\ap#1#2#3{Ann.~of Phys.~{\bf {#1}} (19{#2}) #3} 
\def\prep#1#2#3{Phys.~Rep.~{\bf {#1}} (19{#2}) #3}
\def\jhep#1#2#3{JHEP {\bf {#1}} (19{#2}) #3}
\def\ptp#1#2#3{Prog.~Theor.~Phys.~{\bf {#1}} (19{#2}) #3}

\def\ibid#1#2#3{{\it ibid.}~{\bf {#1}} (19{#2}) #3}

\def\mpl#1#2#3{Mod.~Phys.~Lett.~{\bf A{#1}} (19{#2}) #3}

\def\hepth#1{{hep-th/{#1}}}

\def\texttts#1{\small\texttt{#1}} 
\def\arXive#1{arXiv:{#1}{$\,$}[hep-th]}

\def\pln#1#2#3{Phys.~Lett.~{\bf {#1}B} (20{#2}) #3} 
\def\npn#1#2#3{Nucl.~Phys.~{\bf B{#1}} (20{#2}) #3}

\def\prn#1#2#3{Phys.~Rev.~{\bf D{#1}} (20{#2}) #3}

\def\jhepn#1#2#3{JHEP {\bf {#1}} (20{#2}) #3}
\def\ptpn#1#2#3{Prog.~Theor.~Phys.~{\bf {#1}} (20{#2}) #3}
\def\ijmpn#1#2#3{Int.~Jour.~Mod.~Phys.~{\bf A{#1}} (20{#2}) #3}

\def\<<{<\!\!<} \def\>>{>\!\!>} 
\def\Check#1{{\raise-1.0pt\hbox{\LARGE\v{}}{\hskip -10pt}{#1}}}

\def\eqques{{~\,={\hskip -11.5pt}\raise -1.8pt\hbox{\large ?}
{\hskip 4.5pt}}{}}
\def\fracm#1#2{\,\hbox{\large{${\frac{{#1}}{{#2}}}$}}\,}
\def\fracmm#1#2{\,{{#1}\over{#2}}\,}

\def\frac#1#2{{\textstyle{#1\over\vphantom2\smash{\raise -.20ex
        \hbox{$\scriptstyle{#2}$}}}}}   
\def\sqrttwo{{\sqrt2}}
\def\scst{\scriptstyle}

\def\.{.$\,$}
\def\-{{\hskip 1.5pt}\hbox{-}}

\def\footnotes#1{{\hskip 1pt}\footnotemark$^)$\footnotetext%
{\hsize=6.5in $^)$~{#1}}} 

\def\low#1{\hskip0.01in{\raise -3pt\hbox{${\hskip 1.0pt}\!_{#1}$}}}
\def\low#1{\hskip0.01in{\raise -3pt\hbox{$\!\!\!_{#1}$}}}
\def\ip{{=\!\!\! \mid}}

\begin{document}

\font\tenmib=cmmib10
\font\sevenmib=cmmib10 at 7pt 
\font\fivemib=cmmib10 at 5pt  
\font\tenbsy=cmbsy10
\font\sevenbsy=cmbsy10 at 7pt 
\font\fivebsy=cmbsy10 at 5pt  
\def\BMfont{\textfont0\tenbf \scriptfont0\sevenbf
                              \scriptscriptfont0\fivebf
            \textfont1\tenmib \scriptfont1\sevenmib
                               \scriptscriptfont1\fivemib
            \textfont2\tenbsy \scriptfont2\sevenbsy
                               \scriptscriptfont2\fivebsy}
\def\rlx{\relax\leavevmode}                  
\def\BM#1{\rlx\ifmmode\mathchoice
                      {\hbox{$\BMfont#1$}}
                      {\hbox{$\BMfont#1$}}
                      {\hbox{$\scriptstyle\BMfont#1$}}
                      {\hbox{$\scriptscriptstyle\BMfont#1$}}
                 \else{$\BMfont#1$}\fi}

\font\tenmib=cmmib10
\font\sevenmib=cmmib10 at 7pt 
\font\fivemib=cmmib10 at 5pt  
\font\tenbsy=cmbsy10
\font\sevenbsy=cmbsy10 at 7pt 
\font\fivebsy=cmbsy10 at 5pt  
\def\BMfont{\textfont0\tenbf \scriptfont0\sevenbf
                              \scriptscriptfont0\fivebf
            \textfont1\tenmib \scriptfont1\sevenmib
                               \scriptscriptfont1\fivemib
            \textfont2\tenbsy \scriptfont2\sevenbsy
                               \scriptscriptfont2\fivebsy}
\def\BM#1{\rlx\ifmmode\mathchoice
                      {\hbox{$\BMfont#1$}}
                      {\hbox{$\BMfont#1$}}
                      {\hbox{$\scriptstyle\BMfont#1$}}
                      {\hbox{$\scriptscriptstyle\BMfont#1$}}
                 \else{$\BMfont#1$}\fi}

\def\inbar{\vrule height1.5ex width.4pt depth0pt}
\def\sinbar{\vrule height1ex width.35pt depth0pt}
\def\ssinbar{\vrule height.7ex width.3pt depth0pt}
\font\cmss=cmss10
\font\cmsss=cmss10 at 7pt
\def\ZZ{{}Z {\hskip -6.7pt} Z{}} 
\def\Ik{\rlx{\rm I\kern-.18em k}}  
\def\IC{\rlx\leavevmode
             \ifmmode\mathchoice
                    {\hbox{\kern.33em\inbar\kern-.3em{\rm C}}}
                    {\hbox{\kern.33em\inbar\kern-.3em{\rm C}}}
                    {\hbox{\kern.28em\sinbar\kern-.25em{\rm C}}}
                    {\hbox{\kern.25em\ssinbar\kern-.22em{\rm C}}}
             \else{\hbox{\kern.3em\inbar\kern-.3em{\rm C}}}\fi}
\def\IP{\rlx{\rm I\kern-.18em P}}
\def\IR{\rlx{\rm I\kern-.18em R}}
\def\IN{\rlx{\rm I\kern-.20em N}}
\def\Ione{\rlx{\rm 1\kern-2.7pt l}}
\def\bbbzz{{\Bbb Z}}

%
\def\unredoffs{} \def\redoffs{\voffset=-.31truein\hoffset=-.59truein}
\def\speclscape{\special{ps: landscape}}

\newbox\leftpage \newdimen\fullhsize \newdimen\hstitle\newdimen\hsbody
\tolerance=1000\hfuzz=2pt\def\fontflag{cm}
\catcode`\@=11 
\hsbody=\hsize \hstitle=\hsize 

\def\nolabels{\def\wrlabeL##1{}\def\eqlabeL##1{}\def\reflabeL##1{}}
\def\writelabels{\def\wrlabeL##1{\leavevmode\vadjust{\rlap{\smash%
{\line{{\escapechar=` \hfill\rlap{\sevenrm\hskip.03in\string##1}}}}}}}%
\def\eqlabeL##1{{\escapechar-1\rlap{\sevenrm\hskip.05in\string##1}}}%
\def\reflabeL##1{\noexpand\llap{\noexpand\sevenrm\string\string%
\string##1}}}
\nolabels
%
\global\newcount\secno \global\secno=0
\global\newcount\meqno \global\meqno=1
\def\newsec#1{\global\advance\secno by1\message{(\the\secno. #1)}
\global\subsecno=0\eqnres@t\noindent{\bf\the\secno. #1}
\writetoca{{\secsym} {#1}}\par\nobreak\medskip\nobreak}
\def\eqnres@t{\xdef\secsym{\the\secno.}\global\meqno=1
\bigbreak\bigskip}
\def\sequentialequations{\def\eqnres@t{\bigbreak}}\xdef\secsym{}
\global\newcount\subsecno \global\subsecno=0
\def\subsec#1{\global\advance\subsecno by1%
\message{(\secsym\the\subsecno.%
 #1)}
\ifnum\lastpenalty>9000\else\bigbreak\fi
\noindent{\it\secsym\the\subsecno. #1}\writetoca{\string\quad
{\secsym\the\subsecno.} {#1}}\par\nobreak\medskip\nobreak}
\def\appendix#1#2{\global\meqno=1\global\subsecno=0%
\xdef\secsym{\hbox{#1.}}
\bigbreak\bigskip\noindent{\bf Appendix #1. #2}\message{(#1. #2)}
\writetoca{Appendix {#1.} {#2}}\par\nobreak\medskip\nobreak}
\def\eqnn#1{\xdef #1{(\secsym\the\meqno)}\writedef{#1\leftbracket#1}%
\global\advance\meqno by1\wrlabeL#1}
\def\eqna#1{\xdef #1##1{\hbox{$(\secsym\the\meqno##1)$}}
\writedef{#1\numbersign1\leftbracket#1{\numbersign1}}%
\global\advance\meqno by1\wrlabeL{#1$\{\}$}}
\def\eqn#1#2{\xdef #1{(\secsym\the\meqno)}\writedef{#1\leftbracket#1}%
\global\advance\meqno by1$$#2\eqno#1\eqlabeL#1$$}
%
\newskip\footskip\footskip8pt plus 1pt minus 1pt 
\font\smallcmr=cmr5 
\def\footnotefont{\smallcmr}
\def\f@t#1{\footnotefont #1\@foot}
\def\f@@t{\baselineskip\footskip\bgroup\footnotefont\aftergroup%
\@foot\let\next}
\setbox\strutbox=\hbox{\vrule height9.5pt depth4.5pt width0pt} %
\global\newcount\ftno \global\ftno=0
\def\foot{\global\advance\ftno by1\footnote{$^{\the\ftno}$}}
%
\newwrite\ftfile
\def\footend{\def\foot{\global\advance\ftno by1\chardef\wfile=\ftfile
$^{\the\ftno}$\ifnum\ftno=1\immediate\openout\ftfile=foots.tmp\fi%
\immediate\write\ftfile{\noexpand\smallskip%
\noexpand\item{f\the\ftno:\ }\pctsign}\findarg}%
\def\footatend{\vfill\eject\immediate\closeout\ftfile{\parindent=20pt
\centerline{\bf Footnotes}\nobreak\bigskip\input foots.tmp }}}
\def\footatend{}
\global\newcount\refno \global\refno=1
\newwrite\rfile
%
\def\ref{[\the\refno]\nref}%
\def\nref#1{\xdef#1{[\the\refno]}\writedef{#1\leftbracket#1}%
\ifnum\refno=1\immediate\openout\rfile=refs.tmp\fi%
\global\advance\refno by1\chardef\wfile=\rfile\immediate%
\write\rfile{\noexpand\Item{#1}\reflabeL{#1\hskip.31in}\pctsign}%
\findarg\hskip10.0pt}%
\def\findarg#1#{\begingroup\obeylines\newlinechar=`\^^M\pass@rg}
{\obeylines\gdef\pass@rg#1{\writ@line\relax #1^^M\hbox{}^^M}%
\gdef\writ@line#1^^M{\expandafter\toks0\expandafter{\striprel@x #1}%
\edef\next{\the\toks0}\ifx\next\em@rk\let\next=\endgroup%
\else\ifx\next\empty%
\else\immediate\write\wfile{\the\toks0}%
\fi\let\next=\writ@line\fi\next\relax}}
\def\striprel@x#1{} \def\em@rk{\hbox{}}
\def\lref{\begingroup\obeylines\lr@f}
\def\lr@f#1#2{\gdef#1{\ref#1{#2}}\endgroup\unskip}
\def\semi{;\hfil\break}
\def\addref#1{\immediate\write\rfile{\noexpand\item{}#1}} 
%
\def\listrefs{\footatend\vfill\supereject\immediate\closeout%
\rfile\writestoppt
\baselineskip=14pt\centerline{{\bf References}}%
\bigskip{\frenchspacing%
\parindent=20pt\escapechar=` \input refs.tmp%
\vfill\eject}\nonfrenchspacing}
%
\def\listrefsr{\immediate\closeout\rfile\writestoppt
\baselineskip=14pt\centerline{{\bf References}}%
\bigskip{\frenchspacing%
\parindent=20pt\escapechar=` \input refs.tmp\vfill\eject}%
\nonfrenchspacing}
\def\listrefsrsmall{\immediate\closeout\rfile\writestoppt
\baselineskip=11pt\centerline{{\bf References}} 
\font\smallerfonts=cmr9 \font\it=cmti9 \font\bf=cmbx9%
\bigskip{\smallerfonts{%
\parindent=15pt\escapechar=` \input refs.tmp\vfill\eject}}}
\def\listrefsrmed{\immediate\closeout\rfile\writestoppt
\baselineskip=12.5pt\centerline{{\bf References}}
\font\smallerfonts=cmr10 \font\it=cmti10 \font\bf=cmbx10%
\bigskip{\smallerfonts{%
\parindent=18pt\escapechar=` \input refs.tmp\vfill\eject}}}
\def\startrefs#1{\immediate\openout\rfile=refs.tmp\refno=#1}
\def\xref{\expandafter\xr@f}\def\xr@f[#1]{#1}
\def\refs#1{\count255=1[\r@fs #1{\hbox{}}]}
\def\r@fs#1{\ifx\und@fined#1\message{reflabel %
\string#1 is undefined.}%
\nref#1{need to supply reference \string#1.}\fi%
\vphantom{\hphantom{#1}}\edef\next{#1}\ifx\next\em@rk\def\next{}%
\else\ifx\next#1\ifodd\count255\relax\xref#1\count255=0\fi%
\else#1\count255=1\fi\let\next=\r@fs\fi\next}
\def\figures{\centerline{{\bf Figure Captions}}%
\medskip\parindent=40pt%
\def\fig##1##2{\medskip\item{Fig.~##1.  }##2}}
%

\newwrite\ffile\global\newcount\figno \global\figno=1
\doit0{
\def\fig{fig.~\the\figno\nfig}
\def\nfig#1{\xdef#1{fig.~\the\figno}%
\writedef{#1\leftbracket fig.\noexpand~\the\figno}%
\ifnum\figno=1\immediate\openout\ffile=figs.tmp%
\fi\chardef\wfile=\ffile%
\immediate\write\ffile{\noexpand\medskip\noexpand%
\item{Fig.\ \the\figno. }
\reflabeL{#1\hskip.55in}\pctsign}\global\advance\figno by1\findarg}
\def\listfigs{\vfill\eject\immediate\closeout\ffile{\parindent40pt
\baselineskip14pt\centerline{{\bf Figure Captions}}\nobreak\medskip
\escapechar=` \input figs.tmp\vfill\eject}}
\def\xfig{\expandafter\xf@g}\def\xf@g fig.\penalty\@M\ {}
\def\figs#1{figs.~\f@gs #1{\hbox{}}}
\def\f@gs#1{\edef\next{#1}\ifx\next\em@rk\def\next{}\else
\ifx\next#1\xfig #1\else#1\fi\let\next=\f@gs\fi\next}
}

\newwrite\lfile
{\escapechar-1\xdef\pctsign{\string\%}\xdef\leftbracket{\string\{}
\xdef\rightbracket{\string\}}\xdef\numbersign{\string\#}}
\def\writedefs{\immediate\openout\lfile=labeldefs.tmp %
\def\writedef##1{%
\immediate\write\lfile{\string\def\string##1\rightbracket}}}
\def\writestop{\def\writestoppt%
{\immediate\write\lfile{\string\pageno%
\the\pageno\string\startrefs\leftbracket\the\refno\rightbracket%
\string\def\string\secsym\leftbracket\secsym\rightbracket%
\string\secno\the\secno\string\meqno\the\meqno}%
\immediate\closeout\lfile}}
\def\writestoppt{}\def\writedef#1{}
\def\seclab#1{\xdef #1{\the\secno}\writedef{#1\leftbracket#1}%
\wrlabeL{#1=#1}}
\def\subseclab#1{\xdef #1{\secsym\the\subsecno}%
\writedef{#1\leftbracket#1}\wrlabeL{#1=#1}}
\newwrite\tfile \def\writetoca#1{}
\def\leaderfill{\leaders\hbox to 1em{\hss.\hss}\hfill}
\def\writetoc{\immediate\openout\tfile=toc.tmp
   \def\writetoca##1{{\edef\next{\write\tfile{\noindent ##1
   \string\leaderfill {\noexpand\number\pageno} \par}}\next}}}
\def\listtoc{\centerline{\bf Contents}\nobreak%
 \medskip{\baselineskip=12pt
 \parskip=0pt\catcode`\@=11 \input toc.tex \catcode`\@=12 %
 \bigbreak\bigskip}} 
\catcode`\@=12 
%

\countdef\pageno=0 \pageno=1
\newtoks\headline \headline={\hfil} 
\newtoks\footline 
 \footline={\bigskip\hss\tenrm\folio\hss}
\def\folio{\ifnum\pageno<0 \romannumeral-\pageno \else\number\pageno 
 \fi} 

\def\nopagenumbers{\footline={\hfil}} 
\def\advancepageno{\ifnum\pageno<0 \global\advance\pageno by -1 
 \else\global\advance\pageno by 1 \fi} 
\newif\ifraggedbottom

\def\raggedbottom{\topskip10pt plus60pt \raggedbottomtrue}
\def\normalbottom{\topskip10pt \raggedbottomfalse} 

\def\on#1#2{{\buildrel{\mkern2.5mu#1\mkern-2.5mu}\over{#2}}}
\def\dt#1{\on{\hbox{\bf .}}{#1}}                
\def\Dot#1{\dt{#1}}

\def\eqdot{{\hskip4pt}{\buildrel{\hbox{\LARGE .}} \over =}\,\,{}} 
\def\eqstar{~{\buildrel * \over =}~} 
\def\eqques{~{\buildrel ? \over =}~} 
\def\eqsurface{~{\buildrel^{\,_{_{_\nabla}}} \over =}~} 
\def\eqstartext{{\hskip4pt}{\buildrel * \over =}{}{\hskip4pt}{}}
\def\eqquestext{{\hskip4pt}{\buildrel ? \over =}{}{\hskip4pt}{}} 
\def\eqdottext{{\hskip4.5pt}{\buildrel{\hbox{\LARGE .}} \over =}{}{\hskip3pt}{}} 

\def\lhs{({\rm LHS})} 
\def\rhs{({\rm RHS})} 
\def\lhsof#1{({\rm LHS~of~({#1})})} 
\def\rhsof#1{({\rm RHS~of~({#1})})} 

\def\binomial#1#2{\left(\,{\buildrel 
{\raise4pt\hbox{$\displaystyle{#1}$}}\over 
{\raise-6pt\hbox{$\displaystyle{#2}$}}}\,\right)} 

\def\Dsl{{}D \!\!\!\! /{\,}} 
\def\doubletilde#1{{}{\buildrel{\mkern1mu_\approx\mkern-1mu}%
\over{#1}}{}}

\def\hata{{\hat a}} \def\hatb{{\hat b}} 
\def\hatc{{\hat c}} \def\hatd{{\hat d}} 
\def\hate{{\hat e}} \def\hatf{{\hat f}} 

\def\circnum#1{{\ooalign%
{\hfil\raise-.12ex\hbox{#1}\hfil\crcr\mathhexbox20D}}}

\def\Christoffel#1#2#3{\Big\{ {\raise-2pt\hbox{${\scst #1}$} 
\atop{\raise4pt\hbox{${\scst#2~ #3}$} }} \Big\} }  


 
\font\smallcmr=cmr6 scaled \magstep2 
\font\smallsmallcmr=cmr5 scaled \magstep 1 
\font\largetitle=cmr17 scaled \magstep1 
\font\LargeLarge=cmr17 scaled \magstep5 
\font\largelarge=cmr12 scaled \magstep0

\def\alephnull{\aleph_0}
\def\sqrtoneovertwopi{\frac1{\sqrt{2\pi}}\,} 
\def\twopi{2\pi} 
\def\sqrttwopi{\sqrt{\twopi}} 

\def\rmA{{\rm A}} \def\rmB{{\rm B}} \def\rmC{{\rm C}} 
\def\HatC{\Hat C}

\def\alpr{\a{\hskip 1.2pt}'} 
\def\dim#1{\hbox{dim}\,{#1}} 
\def\leftarrowoverdel{{\buildrel\leftarrow\over\partial}} 
\def\rightarrowoverdel{{\buildrel\rightarrow\over%
\partial}} 
\def\ee{{\hskip 0.6pt}e{\hskip 0.6pt}} 

\def\neq{\not=} 
\def\lowlow#1{\hskip0.01in{\raise -7pt%
\hbox{${\hskip1.0pt} \!_{#1}$}}} 
\def\eqnabla{{~\, }\raise7pt\hbox{${\scriptstyle\nabla}$}{\hskip -11.5pt}={}} 

\def\atmp#1#2#3{Adv.~Theor.~Math.~Phys.~{\bf{#1}}  
(19{#2}) {#3}} 

\font\smallcmr=cmr6 scaled \magstep2 

\def\fracmm#1#2{{{#1}\over{#2}}} 
\def\fracms#1#2{{{\small{#1}}\over{\small{#2}}}} 
\def\low#1{{\raise -3pt\hbox{${\hskip 1.0pt}\!_{#1}$}}} 
\def\medlow#1{{\raise -1.5pt\hbox{${\hskip 1.0pt}\!_{#1}$}}}

\def\mplanck{M\low{\rm P}} 
\def\mplancktwo{M_{\rm P}^2} 
\def\mplanckthree{M_{\rm P}^3} 
\def\mplanckfour{M_{\rm P}^4} 
\def\mweylon{M\low S}  
\def\mhiggs{M_\medlow H}
\def\mwboson{M \low{\rm W}} 

\def\ip{{=\!\!\! \mid}} 
\def\Lslash{${\rm L}{\!\!\!\! /}\, $} 

\def\leapprox{~\raise 3pt \hbox{$<$} \hskip-9pt \raise -3pt \hbox{$\sim$}~} 
\def\geapprox{~\raise 3pt \hbox{$>$} \hskip-9pt \raise -3pt \hbox{$\sim$}~} 

\def\fR{f (R ) }
\def\FR{F \[ R \]} 
\def\FLaginv{F \[ e^{-1} \Lag_{\rm inv} \]}  
\def\LagSG{\Lag_{\rm SG}} 
\def\Laginv{\Lag_{\rm inv}} 
\def\Lagtot{\Lag_{\rm tot}} 
\def\FprimeLaginv{F\, ' \[e^{-1} \Lag_{\rm inv} \] }   
\def\FdoubleprimeLaginv{F\, '' \[e^{-1} \Lag_{\rm inv} \] }  
\def\Fzeroprime{F\, '\!\!\!_0\,}

\def\qed{(\hbox{\it Q.E.D.})}

\def\sqrttwo{{\sqrt 2}} 

\def\squarebrackets#1{\left[ \, {#1} \, \right]}  



\def\gswcont{M.B.~Green, J.H.~Schwarz and E.~Witten, 
{\it `Superstring Theory'}, Vols.~I \& II, 
Cambridge Univ.~Press (1986).} 

\def\mtheorycont{C.~Hull and P.K.~Townsend,
\np{438}{95}{109}; E.~Witten, \np{443}{95}{85}; 
P.K.~Townsend, {\it `Four Lectures on M-Theory'}, in {\it
`Proceedings of ICTP Summer School on High Energy
Physics and Cosmology'}, Trieste (June 1996),
hep-th/9612121;  {\it `M-theory from its Superalgebra'}, Cargese Lectures, 
1997, hep-th/9712004; T.~Banks, W.~Fischler, S.H.~Shenker
and L.~Susskind, \pr{55}{97}{5112}; 
K.~Becker, M.~Becker and J.H.~Schwarz, 
{\it `String Theory and M-Theory:  A Modern Introduction'}, 
Cambridge University Press, 2007.} 

\def\dbicont{M.~Born and L.~Infeld, Proc.~Roy.~Soc.~Lond.~%
{\bf A143} (1934) 410; {\it ibid.}~{\bf A144} (1934) 425;
P.A.M.~Dirac, Proc.~Roy.~Soc.~Lond.~{\bf A268} (1962) 57.}  

\def\gzcont{M.K.~Gaillard and B.~Zumino, 
\np{193}{81}{221}.}  

\def\bracecont{G.W.~Gibbons and D.A.~Rasheed
\np{454}{95}{185}, \hepth{9506035};  
D.~Brace, B.~Morariu and B.~Zumino, 
In *Shifman, M.A.~(ed.)~{\it `The many faces of the superworld'}, 
pp.~103-110, \hepth{9905218}; 
M.~Hatsuda, K.~Kamimura and S.~Sekiya, 
Nucl.~Phys.~{\bf 561} (1999) 341; 
P.~Aschieri, \ijmpn{14}{00}{2287}.}  

\def\kuzenkocont{S.~Kuzenko and S.~Theisen, 
\jhepn{03}{00}{034}.}   

\def\pvncont{P.~van Nieuwenhuizen, \prep{68}{81}{189}.}  

\def\schwarzsencont{J.H.~Schwarz and A.~Sen, \np{411}{94}{35}, 
\hepth{9304154}.}  

\def\shapereetalcont{A.D.~Shapere, S.~Trivedi and F.~Wilczek, 
\mpl{6}{91}{2677};
A.~Sen, 
\np{404}{93}{109}.}  
 
\def\aschierietalcont{
P.~Aschieri, D.~Brace, B.~Morariu and B.~Zumino
\npn{574}{00}{551}, \hepth{9909021}.} 
 
\def\nrtencont{H.~Nishino and S.~Rajpoot, 
\prn{71}{05}{085011}.} 

\def\sevenformcont{H.~Nicolai, P.K.~Townsend and P.~van Nieuwenhuizen, 
Lett.~Nuov.~Cim.~{\bf 30} (1981) 315; 
R.~D'Auria and P. Fr\' e, \np{201}{82}{101}.}  

\def\branecont{P.K.~Townsend, 
{\it `p-Brane Democracy'}, hep-th/9507048; 
H.~Nishino, 
\mpl{14}{99}{977}, \hepth{9802009}.}  

\def\bbscont{I.A.~Bandos, N.~Berkovits and D.P.~Sorokin, 
\np{522}{98}{214}, \hepth{9711055}.}

\def\dwscont{B.~de Wit and H.~Samtleben, 
Fortsch.~Phys.~{\bf 53} (2005) 442, hep-th/0501243.}

\def\dwnscont{B.~de Wit, H.~Nicolai and H.~Samtleben, 
JHEP 0802:044,2008, \arXive{0801.1294}.}  

\def\chucont{C.-S.~Chu, 
{\it `A Theory of Non-Abelian Tensor Gauge Field
with Non-Abelian Gauge Symmetry $G \times G$'}, 
\arXive{1108.5131}.}  

\def\sswcont{H.~Samtleben, E.~Sezgin and R.~Wimmer, 
\jhepn{12}{11}{062}.}  

\def\nrthreecont{H.~Nishino and S.~Rajpoot, 
\prn{82}{10}{087701}.}  

\def\ntcont{H.~Nicolai and P.K.~Townsend, \pl{98}{81}{257}.}   

\def\englertwindeycont{F.~Englert and P.~Windey, 
\pr{14}{76}{2728}.} 

\def\montonenolivecont{C.~Montonen and D.I.~Olive, 
\pl{72}{77}{117}.} 

\def\olivewittencont{D.I.~Olive and E.~Witten,
\pl{78}{78}{97}.} 

\def\osborncont{H.~Osborn, 
\pl{83}{79}{321}.} 

\def\cjscont{E.~Cremmer, B.~Julia and J.~Scherk, \pl{76}{78}{409};
E.~Cremmer and B.~Julia, \pl{80}{78}{48}; \np{159}{79}{141}.}

\def\pstcont{P.~Pasti, D.P. Sorokin, M.~Tonin, 
\pl{352}{95}{59}, \hepth{9503182}.}     

\def\htwcont{C.~Hull, P.K.~Townsend, \np{438}{95}{109}; 
E.~Witten, \np{443}{95}{85}.}  

\def\tseytlinetalcont{A.A.~Tseytlin, \np{469 }{96}{51}; 
Y.~Igarashi, K.~Itoh and K.~Kamimura, \np{536 }{99}{469}.}  

\def\sstcont{M.B.~Green, J.H.~Schwarz and E.~Witten, 
{\it `Superstring Theory'}, Vols.~I \& II, 
Cambridge Univ.~Press (1986); 
K.~Becker, M.~Becker and J.H.~Schwarz, 
{\it `String Theory and M-Theory:  A Modern Introduction'}, 
Cambridge University Press, 2007.} 

\def\nogocont{M.~Henneaux, V.E.~Lemes, C.A.~Sasaki, S.P.~Sorella, 
O.S.~Ventura and L.C.~Vilar, \pl{410}{97}{195}.}  

\def\ftcont{D.Z.~Freedman, P.K.~Townsend, \np{177}{81}{282};
{\it See also}, V.I.~Ogievetsky and I.V.~Polubarinov, 
Sov.~J.~Nucl.~Phys.~{\bf 4} (1967) 156} 

\def\ksmcont{S.~Krishna, A.~Shukla and R.P.~Malik, 
\ijmpn{26}{11}{4419}, \arXive{1008.2649}.} 

\def\topologicalcont{J.~Thierry-Mieg and L.~Baulieu, \np{228}{83}{259}; 
A.H.~Diaz, \pl{203}{88}{408}; 
T.J.~Allen, M.J.~Bowick and A.~Lahiri, \mpl{6}{91}{559};
A.~Lahiri, \pr{55}{97}{5045};
E.~Harikumar, A.~Lahiri and M.~Sivakumar, \prn{63}{01}{10520}.}

\def\ferraracont{G.~Dall'Agata and S.~Ferrara, 
\npn{717}{05}{223}, \hepth{0502066}; 
G.~Dall'Agata, R.~D'Auria and S.~Ferrara, 
\pln{619}{05}{149}, \hepth{0503122}; 
R.~D'Auria and S.~Ferrara, 
\npn{732}{06}{389}, \hepth{0504108}; 
R.~D'Auria, S.~Ferrara and M.~Trigiante, 
\jhep{0509}{05}{035}, \hepth{0507225}.}  

\def\nrnacont{H.~Nishino and S.~Rajpoot, 
\hepth{0508076}, \prn{72}{05}{085020}.} 

\def\finncont{U.~Lindstrom and M.~Ro\v cek, \np{222}{85}{285};
T.E.~Clark, C.H.~Lee and S.T.~Love, \mpl{4}{89}{1343};
F.~Brandt and U.~Theis, \np{550}{99}{495}; 
K.~Furuta, T.~Inami, H.~Nakajima and M.~Nitta, 
\ptpn{106}{01}{851}, \hepth{0106183}.} 
 
\def\scherkschwarzcont{J.~Scherk and J.H.~Schwarz, \np{153}{79}{61}.}     

\def\nepomechiecont{R.I.~Nepomechie, \np{212}{83}{301}.} 

\def\problemnonabeliancont{J.M.~Kunimasa and T. Goto, \ptp{37}{67}{452}; 
A.A.~Slavnov, Theor.~Math.~Phys.~{\bf 10} (1972) 99;
M.J.G.~Veltman, \np{7}{68}{637}; 
A.A.~Slavnov and L.D.~Faddeev, Theor.~Math.
\newline Phys.~{\bf 3} (1970) 312; 
A.I.~Vainshtein and I.B.~Khriplovich, Yad.~Fiz.~13 (1971) 198; 
K.I.~Shizuya, \np{121}{77}{125}; 
Y.N.~Kafiev, \np{201}{82}{341}.    
}

\def\aflcont{L.~Andrianopoli, S.~Ferrara and M.A.~Lledo, 
\hepth{0402142}, \jhep{0404}{04}{005};  
R.~D'Auria, S.~Ferrara, M.~Trigiante and S.~Vaula, 
\pln{610}{05}{270}, \hepth{0412063}.}   

\def\stueckelbergcont{E.C.G.~Stueckelberg, Helv.~Phys.~Acta {\bf 11}  (1938) 225;
A.~Proca, J.~Phys.~Radium {\bf 7} (1936) 347; 
{\it See, e.g.}, R.~Delbourgo and G.~Thompson, \prl{57}{86}{2610}
D.~Feldman, Z.~Liu and P.~Nath, \prl{97}{86}{021801}; 
{\it For reviews, see, e.g.}, H.~Ruegg and M.~Ruiz-Altaba, \ijmpn{19}{04}{3265}.} 

\def\mt{M.~Blau and G.~Thompson,
\ap{205}{91}{130}.}  

\def\hk{M.~Henneaux and B.~Knaepen, 
\pr{56}{97}{6076}, \hepth{9706119}.}

\def\originalcont{S.~Ferrara, B.~Zumino, and J.~Wess, \pl{51}{74}{239}; 
W.~Siegel, \pl{85}{79}{333}; U.~Lindstrom and M.~Ro\v cek, 
\np{222}{83}{285}; 
{\it For reviews of linear multiplet coupled to SG, see, e.g}., P.~Bine«truy, 
G.~Girardi, and R.~Grimm, \prn{343}{01}{255}, {\it and
references therein}.} 

\def\stringrelatedcont{S.~Ferrara and M.~Villasante, \pl{186}{87}{85};
P.~Bin\' etruy, G.~Girardi, R.~Grimm, and M.~Muller, \pl{195}{87}{389}; 
S.~Cecotti, S.~Ferrara, and M.~Villasante, Int.~Jour.~Mod.~Phys.
\newline {\bf A2} (1987) 1839; 
M.K.~Gaillard and T.R.~Taylor, \np{381}{92}{577};
V.S.~Kaplunovsky and J.~Louis, \np{444}{95}{191};
P.~Bin\' etruy, F.~Pillon, G.~Girardi and R.~Grimm, \np{477}{96}{175}; 
P.~Bin\' etruy, M.K.~Gaillard and Y.-Y.~Wu, \pl{412}{97}{288}; 
\np{493}{97}{27}493, \ibid{B481}{96}{109};
D.~Lu¬st, S.~Theisen and G.~Zoupanos, \np{296}{88}{800}; 
J.~Lauer, D.~L\" ust and S.~Theisen, \np{304}{88}{236}.}  

\def\threealgebracont{N.~Lambert and C.~Papageorgakis, 
\jhepn{08}{10}{083}; 
K-W.~Huang and W-H.~Huang, 
arXiv:{1008.3834} \newline [hep-th]; 
S.~Kawamoto, T.~Takimi and D.~Tomino, 
J.~Phys.~{\bf A44} (2011) 325402, \arXive{1103.1223};
Y.~Honma, M.~Ogawa, S.~Shiba, 
\jhepn{1104}{11}{2011}, \arXive{1103.1327}; 
C.~Papageorgakis and C.~Saemann, \jhep{1105}{11}{099}, 
\arXive{1103.6192}.}  

\def\wbcont{J.~Wess and J.~Bagger, {\it `Superspace and Supergravity'}, 
Princeton University Press (1992).}  
 

\def\sgcont{S.~Ferrara, D.Z.~Freedman and P.~van Nieuwenhuizen, 
\pr{13}{76}{3214};
S.~Deser and B.~Zumino, \pl{62}{76}{335}; 
P.~van Nieuwenhuizen, \prep{68}{81}{189};
J.~Wess and J.~Bagger, {\it `Superspace and Supergravity'}, 
Princeton University Press (1992).}  
 
\def\typeiibcont{J.H.~Schwarz, \np{226}{83}{269}.} 



\def\framing#1{\doit{#1}  {\framingfonts{#1} 
\border\headpic  }} 

\framing{0} 

\def\Cases#1{\left \{ \matrix{\displaystyle #1} \right.}   

\def\fIJK{f^{I J K}} 
\def\eqquestext{{\hskip4pt}{\buildrel ? \over =}{}{\hskip4pt}{}} 

\doit0{
\def\matrix#1{\null\ , \vcenter{\normalbaselines\m@th
	\ialign{\hfil$##$\hfil&&\quad\hfil$##$\hfil\crcr 
	  \mathstrut\crcr\noalign{\kern-\baselineskip}
	  #1\crcr\mathstrut\crcr\noalign{\kern-\baselineskip}}}\ ,} 
} 

\def\ialign{\everycr={}\tabskip=0pt \halign} 

\doit0{
\def\matrixs#1{\null\ , {\normalbaselines \m@th
	\ialign{\hfil$##$\hfil&&\quad\hfil$##$\hfil\crcr 
	  \mathstrut\crcr\noalign{\kern-\baselineskip}
	  #1\crcr\mathstrut\crcr\noalign{\kern-\baselineskip}}}\ ,} 
} 



\doit0{
\vskip -0.6in 
{\bf Preliminary Version (FOR YOUR EYES
ONLY!)\hfill\today} \\[-0.25in] 
\\[-0.3in]  
}
\vskip -0.3in  
\doit0{
{\hbox to\hsize{\hfill
hep-th/yymmnnn}} 
}
\doit1{\vskip 0.1in 
{\hbox to\hsize{\hfill CSULB--PA--11--3}} 
\vskip -0.05in 
{\hbox to\hsize{\hfill (Revised Version)}} 
} 

\vskip 0.55in 

\begin{center} 

{\Large\bf N$\,$=$\,$1$\,$ Non-Abelian Tensor Multiplet} \\
\vskip 0.05in 
{\Large\bf in Four Dimensions} \\ 
\vskip 0.1in 

\baselineskip 9pt 

\vskip 0.21in 

Hitoshi ~N{\smallcmr ISHINO}%
\footnotes{E-Mail: hnishino@csulb.edu} ~and
~Subhash ~R{\smallcmr AJPOOT}%
\footnotes{E-Mail: rajpoot@csulb.edu} 
\\[.16in]  {\it Department of Physics \& Astronomy}
\\ [.015in] 
{\it California State University} \\ [.015in]  
{\it 1250 Bellflower Boulevard} \\ [.015in]  
{\it Long Beach, CA 90840} \\ [0.02in] 

\vskip 1.8in 

{\bf Abstract}\\[.1in]  
\end{center} 
\vskip 0.1in 

\baselineskip 14pt

~~~We carry out the $~N=1$~ supersymmetrization of a physical 
non-Abelian tensor with non-trivial consistent couplings in four dimensions.  
Our system has three multiplets:  
(i) The usual non-Abelian vector multiplet (VM) $~(A\du\m I, \l^I)$, ~(ii) 
A non-Abelian tensor multiplet (TM) $~(B\du{\m\n} I , \chi^I, \varphi^I)$, 
and (iii) A compensator vector multiplet (CVM) $\,(C\du\m I, \r^I)$.  All of these 
multiplets are in the adjoint representation of a non-Abelian group $\, G$.  
Unlike topological theory, all of our fields are propagating with kinetic terms.  
The $C\du\m I\-$field plays the role of a Stueckelberg 
compensator absorbed into the longitudinal component of $~B\du{\m\n} I$.  
We give not only the component lagrangian, but also a corresponding 
superspace reformulation, reconfirming the total consistency of the system.  
The adjoint representation of the TM and CVM is further generalized to 
an arbitrary real representation of general $~SO(N)$~ gauge group.  We also couple the globally $~N=1$~ supersymmetric system to supergravity, 
as an additional non-trivial confirmation.

\vskip 0.5in  

\baselineskip 8pt 
\leftline{\small PACS:  11.15.-q, 11.30.Pb, 12.60.Jv}  
\vskip 0.03in 
\leftline{\small Key \hfil Words:  \hfil Non-Abelian Tensor, 
\hfil $~N=1$~ Supersymmetry, \hfill Tensor Multiplet, \hfill Vector Field} 
\leftline{\small {\hskip 0.8in} \hfil in Non-Trivial Representation, 
\hfil Consistency of Field Equations and Couplings.}


\vfill\eject  

\baselineskip 18pt 

\oddsidemargin=0.03in 
\evensidemargin=0.01in 
\hsize=6.5in
\topskip 0.32in 
\textwidth=6.5in 
\textheight=9in 
\flushbottom
\footnotesep=1.0em
\footskip=0.36in 
\def\baselinestretch{1.0} 

\def\fixedpoint{19.0pt} 
\baselineskip\fixedpoint    

\pageno=2 



\leftline{\bf 1.~~Introduction}  

Recently, the long-standing problem with non-Abelian tensors
\ref\nogo{\nogocont}  
has been solved by de Wit, Samtleben, and Nicolai
\ref\dws{\dwscont}%
\ref\dwns{\dwnscont}.   
The original motivation in \dws\ was to generalize the tensor and vector field 
interactions in manifestly $~E_{6(+6)}\-$covariant formulation of 
five-dimensional (5D) maximal supergravity by gauging non-Abelian sub-groups.  In \dwns, this work was further related to M-theory 
\ref\mtheory{\mtheorycont} 
by confirming the representation assignments under the duality group of the gauge charges.  The underlying hierarchies of these tensor and vector gauge fields 
are presented with the consistency of general gaugings.  

The hierarchy in \dws\dwns\ has been further applied to the 
conformal supergravity in 6D 
\ref\ssw{\sswcont}.
In ref.~\ssw, the `minimal tensor hierarchy' as a special case of the more general 
hierarchy in \dws\dwns\ has been discussed.  
This hierarchy consists of $~A\du\m r $~ 
and two-form gauge potentials $B\du{\m\n} I$, 
with two labels $~{\scst r}$~ and $~{\scst I}$.  Also introduced is 
the 3-form gauge potentials $~C_{\m\n\r\, r}$~ with the index $~_r$~ is dual to $~^r$~ of $~A\du\m r$.  
The field strengths of vector and two-form gauge potentials are defined by \ssw
$$ \li{\calF\du{\m\n} r & \equiv 2 \partial_{\[ \m } A\du{\n \]} r 
		+ h\du I r B\du{\m\n} I ~~, 
&(1.1\rma) \cr 
\calH\du{\m\n\r} I &  \equiv 3 D_{\[\m} B\du{\n\r\]} I 
				+ 6 d\du{r s} I A\du{\[ \m} r \partial_\n A\du{\r\]} s 
				- 2 f\du{p q} s d\du{r s} I A\du{\[\m} r A\du\n p A\du{\r\]} q 
				+ g^{I r} C_{\m\n\r r} ~~. ~~~~~ 
&(1.1\rmb) \cr } $$ 

The prescription for tensor-vector system, 
which we will be based upon, is described with eq.$\,$(3.22) in  
\ssw.  To be more specific, 
we consider in the present paper the product of two identical gauge groups 
$~G \times G$~ 
\ref\chu{\chucont},  
whose adjoint indices are 
respectively $~{\scst r,~s,~\cdots}$~ and $~{\scst r', ~s',~\cdots}$.      
Accordingly, we use the coefficients 
$$ \li{ f\du{r s} t & = {\rm f}\du{r s} t ~~, ~~~~ f\du{r s'} {t'} = - f\du{s' r} {t'} 
					= +\frac 12 {\rm f}\du{r s'} {t'} ~~, 
&(1.2\rma) \cr  
d_{r s'}^t & = d_{s' r}^t  = - \frac 12 {\rm f}\du{r s'} t ~~, ~~~~ 
				h^{r'}_s = \d_s ^{r'} ~~,  
&(1.2\rmb) \cr } $$  
where $~{\rm f}\du{r s} t $~ is the structure constant of a non-Abelian gauge group.  
We use the same field content arising by this prescription.  

\doit0{
The notational correspondence between \ssw\ and ours in the present paper 
is summarized as 
$$\li{ A\du\m r ~~~~ & \Longleftrightarrow ~~~~~ A\du\m I ~~~~, 
&(1.3\rma) \cr
B\du{\m\n} I  ~~~~ & \Longleftrightarrow ~~~~~ B\du{\m\n} I ~~~~, 
&(1.3\rmb) \cr
A\du\m {r'} ~~~~ & \Longleftrightarrow ~~~~~ C\du\m I ~~.  
&(1.3\rmc) \cr } $$  
Accordingly, the field strengths of our potentials $~B\du{\m\n} I$~ 
and $~C\du\m I$~ are  
$$\li{ G\du{\m\n\r} I & \equiv + 3 D_{\[\m} B\du{\n\r\] } I 
		- 3 f^{I J K} C\du{\[\m} J F\du{\n\r\] } K  ~~, 
&(1.4\rma) \cr 
H\du{\m\n} I & \equiv + 2 D_{\[\m } C\du{\n\]} I + g B \du{\m\n} I ~~.
&(1.4\rmb) \cr } $$ 
The mass dimension of the bosonic fields $~A,~B$~ 
and $~C\-$fields~ is zero, while that of the coupling constant $~g$~ 
to be unity, following the convention in superspace.  The last term in (1.4b) 
indicates that the vector $~C\du\m I$~ is nothing but a Stueckelberg
\ref\stueckelberg{\stueckelbergcont}
type vector, as in the original hierarchy \dws\dwns\ssw.   
}

Since the outstanding paper \ssw\ gives the extensive details of 
how to get our system from \dws\dwns\chu, there is nothing new to explain, 
except for our notational preparation.   
In our notation, the field strengths of the $~B$~ and $~C\-$fields 
are respectively $~G$~ and $~H$~ defined by 
$$\li{ G\du{\m\n\r} I & \equiv + 3 D_{\[\m} B\du{\n\r\] } I 
		- 3 f^{I J K} C\du{\[\m} J F\du{\n\r\] } K  ~~, 
&(1.3\rma) \cr 
H\du{\m\n} I & \equiv + 2 D_{\[\m } C\du{\n\]} I + g B\du{\m\n} I ~~.
&(1.3\rmb) \cr } $$ 
The gauge transformations for $~B,~C$~ and $~A\-$fields are  
$$\li{ \d_\a (B\du{\m\n} I , C\du\m I, A\du\m I ) 
& = ( \, - \fIJK \a^J B\du{\m\n} K , ~- \fIJK \a^J C\du\m K ,~ + D_\m \a^I ) ~~, 
&(1.4\rma) \cr 
\d_\b ( B\du{\m\n}I,  C\du\m I ,A\du\m I ) 
& = ( \, + 2 D_{\[\m} \b\du{\n\]}  I , ~ - g \b\du\m I , ~ 0) ~~, 
&(1.4\rmb) \cr 
\d_\g ( B\du{\m\n} I, C\du\m I ,A\du\m I ) 
& = ( \, - \fIJK F\du{\m\n} J \g^K , ~D_\m \g^I , ~0)  ~~. 
&(1.4\rmc) \cr} $$ 
As (1.3b) or (1.4b) shows, $~C\du\m I$~ is a vectorial Stueckelberg field, 
absorbed into the longitudinal component of $~B\du{\m\n} I$.   
Due to the general hierarchy \dws\dwns, all field strengths are invariant: 
$$\li{ & \d_\a (G\du{\m\n\r} I , ~H\du{\m\n} I, ~F\du{\m\n} I ) 
	= - \fIJK \a^J ( G\du{\m\n\r} K ,~ H\du{\m\n} K , ~F\du{\m\n} K ) ~~, 
&(1.5\rma) \cr 
& \d_\b (G\du{\m\n\r} I , ~H\du{\m\n} I , ~F\du{\m\n} I)  = 0 ~~,
		~~~~ \d_\g (G\du{\m\n\r} I , ~H\du{\m\n} I ,~ F\du{\m\n} I)  = 0 ~~.   
&(1.5\rmb) \cr} $$ 

Since the hierarchy given in \dws\dwns\ guarantees the gauge invariance of 
all field strengths, the construction of purely bosonic lagrangian is 
straightforward.  Consider the action $~I_1 \equiv \int d^4 x \, g^2 \Lag_1$
\footnotes{The reason we need the factor $~g^2$~ in the action is due to 
the mass-dimension assignments of our fields.} with  
$$\li{ & \Lag_1 \equiv -\frac 1{12} ( G\du{\m\n\r} I )^2
		 - \frac 14 ( H\du{\m\n} I)^2 - \frac 14 ( F\du{\m\n} I)^2 ~~. 
&(1.6) \cr } $$  
The gauge invariances of all field strength also guarantee the 
consistency of the $~A,~B$~ and $~C\-$field equations, 
such as the divergence $~D_\n (\d \Lag_1 / \d B\du{\m\n} I ) \eqdot 0$.\footnotes{We use the symbol $~\eqdot$~ for a field equation to be distinguished from an 
algebraic equation.}  Since we will do similar confirmation for supersymmetric 
system later, we skip the details for the purely bosonic system.  

The purpose of our present paper is to supersymmetrize this system.  
The rest of our paper is organized as follows.  
In section 2, we give the component formulation of 
$~N=1$~ tensor multiplet (TM).  
In section 3, we give the superspace re-formulation of 
component result.  In section 4, we give the generalization to non-adjoint 
representation of $~G=SO(N)$~ case.  In section 5, we give the 
supergravity coupling to non-Abelian TM, as supporting evidence for
the consistency of the global case.  Section 6 is for concluding 
remarks.  Appendix A is devoted to purely bosonic systems of 
non-Abelian tensors with much simpler structures than has been 
presented in arbitrary space-time dimensions with arbitrary signature.  
An example of tensor-vector duality $~G = F^*$~ in $~D=2+4$~ dimensions, 
and its dimensional reduction (DR) into the self-dual YM $~F = F^*$~ 
in $~D=2+2$~ is also presented.  

\bigskip\bigskip 




\leftline{\bf 2.~~Component Formulation of $\,$N=1$\,$ TM} 
\nobreak 

The supersymmetrization of the purely bosonic system (1.6) is 
rather straightforward, except for subtlety to be mentioned later.    
Our system has three multiplets: (i) A TM $(B\du{\m\n} I, \chi^I, \varphi^I)$, 
(ii) A compensating vector multiplet (CVM) $(C\du\m I, \r^I)$, and (iii) 
A Yang-Mills vector multiplet (YMVM) $(A\du\m I, \l^I)$.  
Our total action $~I \equiv \int d^4 x\, \Lag$~ has the lagrangian 
$$\li{ \Lag = & - \frac 1{12} ( G\du{\m\n\r} I)^2
			 + \frac 12 (\Bar\chi{}^I \Dsl\chi^I) 
			- \frac 12 (D_\m\varphi^I)^2  
	- \frac 12 g^2 ( \varphi^I)^2 - g (\Bar\chi^I \r^I) \cr 
& - \frac 14 (H\du{\m\n} I)^2 + \frac 12 (\Bar\r{}^I \Dsl\r^I) 
   	 - \frac 14 (F\du{\m\n} I)^2 + \frac 12 (\Bar\l{}^I \Dsl\l^I) \cr  
& -\frac12 g f^{I J K} (\Bar\l{}^I \chi^J) \varphi^K  
		+\frac12 f^{I J K} (\Bar\l^I \g^\m \r^J) D_\m \varphi^K 
		+\frac1{12} f^{I J K} (\Bar\l^I \g^{\m\n\r} \r^J ) G\du{\m\n\r} K  
		~~~~~ ~~\cr  
& + \frac 14 f^{I J K} (\Bar\r{}^I \g^{\m\n} \chi^J) F\du{\m\n} K 
		-\frac 14 f^{I J K} (\Bar\l^I \g^{\m\n} \chi^J) H_{\m\n}{}^K  
		-\frac 12 f^{I J K} F\du{\m\n} I H^{\m\n\, J} \varphi^K ~~, 
& (2.1) \cr} $$  
up to quartic-order terms $~\order\phi4$.  

It is clear that the scalar $~\varphi^I$~ has its mass $~g$, while there is a 
mixture between $~\chi^I$~ and $~\r^I$, again with the asme mass $~g$.  
As has been mentioned after (1.4), $~C\du\m I $~ plays the role of 
Stueckelberg field 
\ref\stueckelberg{\stueckelbergcont}, 
being absorbed into the longitudinal component of $~B\du{\m\n} I$.  
Eventually, the kinetic term of the $~C\-$field becomes the mass term of 
$~B\du{\m\n} I$.  Accordingly, the degrees of freedom (DOF) for 
the massive TM fields are $~B\du{\m\n} I ~(3), ~\r^I ~(4)$~ and 
$~\varphi^I (1)$, up to the adjoint index $~{\scst I}$.     

Our action $~I$~ is invariant under global $~N=1$~ supersymmetry 
$$\li{ \d_Q B\du{\m\n} I = &  + (\Bar\e\g_{\m\n} \chi^I) 
					- 2 \fIJK C\du{\[\m | } J (\d_Q A\du{| \n\]} K) ~~, 
&(2.2\rma) \cr  
\d_Q \chi^I = & + \frac 16 (\g^{\m\n\r} \e) G_{\m\n\r}{}^I 
				- (\g^\m\e) D_\m \varphi^I \cr 
& + \frac 12 \fIJK \Big[ + \e (\Bar\l{}^J\r^K) 
					- (\g_5\g^\m \e) (\Bar\l{}^J \g_5\g_\m \r^K) 
					  -  (\g_5 \e) (\Bar\l{}^J \g_5\r^K)\Big] ~~, 
&(2.2\rmb) \cr  
\d_Q \varphi^I = & + (\Bar\e\chi^I) ~~, 
&(2.2\rmc) \cr  
\d_Q C\du\m I =  & +(\Bar\e\g_\m\r^I) 
		+ f^{I J K} (\Bar\e \g_\m \l^J) \varphi^K ~~, 
&(2.2\rmd) \cr  
\d_Q \r^I =  & + \frac 12 (\g^{\m\n}\e) H\du{\m\n} I - g \e \varphi^I  
		- \frac 12 f^{I J K} (\g^{\m\n}\e) F\du{\m\n} J\varphi^K \cr 
& + \frac 14 f^{I J K} \Big[ + \e (\Bar\l{}^J \chi^K) 
			- (\g^\m\e) (\Bar\l{}^J \g_\m\chi^K ) 
			+ \frac 12 (\g^{\m\n}\e) (\Bar\l{}^J \g_{\m\n} \chi^K )  \cr 
& ~~~~~ ~~~~~ ~~\,\, - (\g_5\g^\m\e) (\Bar\l{}^J \g_5 \g_\m\chi^K ) 
			- (\g_5 \e) (\Bar\l{}^J \g_5\chi^K ) \Big] ~~, 
&(2.2\rme) \cr
\d_Q A\du\m I =  & + (\Bar\e\g_\m\l^I) ~~, 
&(2.2\rmf) \cr 
\d_Q \l^I = & + \frac12 (\g^{\m\n}\e) F\du{\m\n} I 
			+ \frac 12 \fIJK (\g_5\e) (\Bar\r^J\g_5 \chi^K) ~~,    
&(2.2\rmg) \cr } $$  
up to cubic terms $~\order\phi 3$~ in fields.   
The fermionic quadratic terms in (2.2b), (2.2e) and (2.2g) are fixed 
in superspace formulation, as will be explained later.  
In the {\it conventional} dimensions 
with all the bosonic (or fermionic) fields with $~1$~ (or $~3/2$) mass 
dimensions,\footnotes{Our bosonic (or fermionic) fields have dimensions 
$~0$ (or $~1/2$), in contrast to the conventional dimensions  
$~1$ (or $~3/2$).} 
these terms lead to non-renormalizability.  For example, the l.h.s.~of (2.2b) has dimension ~$3/2$, while its r.h.s.~for the $~\e (\Bar\l \g\r)$~
term has $~(-1/2) + (3/2) + (3/2) = 5/2$.  In other words, there is an implicit 
coupling constant $~\ell$~ with the dimension of length in front of 
fermionic quadratic terms.  This feature is also related to the existence of Pauli-terms which are non-renormalizable, already at a {\it globally} supersymmetric 
system.  These features are similar to supergravity 
\ref\sg{\sgcont},  
even though our system so far has only {\it global} supersymmetry.   

The usual non-Abelian gauge transformation $~\d_\a$~ and our tensorial 
gauge transformation $~\d_\b$, and $~\d_\g\-$transformation are 
exactly the same as (1.4),  
while all the fermionic fields are transforming only under $~\d_\a$, as 
the $~B$~ and $~C\-$fields do,     
so that there arises no problem with the $~\d_\b$~ and $~\d_\g\-$invariances 
of the field strengths as in (1.5).  
These immediately lead to the invariances of our action $~\d_\a I =0, ~
\d_\b I = 0$~ and $~\d_\g I = 0$.  

The Bianchi identities (BIds) for our field strengths $~G, ~H$~ and $~F$~ are:
$$ \li{ D_{\[\m} G\du{\n\r\s\]} I 
		 - \fracm 3 2 f^{I J K} F\du{\[\m\n} J H\du{\r\s\]} K & \equiv 0 ~~, 
&(2.3\rma) \cr 
D_{\[ \m} H\du{ \n\r\]} I - \fracm 1 3 g\, G_{\m\n\r}{}^I & \equiv 0 ~~, 
&(2.3\rmb)  \cr 
D_{\[ \m} F\du{ \n\r\]}I & \equiv 0 ~~.   
&(2.3\rmc)  \cr } $$ 
Relevantly, the non-trivial $~\d_Q\-$transformations 
of the field strengths are 
$$ \li{ \d_ Q G\du{\m\n\r} I = & + 3(\Bar\e \g_{\[\m\n} D_{\r\]} \chi^I) 
			+ 3 \fIJK (\d_Q A\du{\[\m } J) H\du{\n\r\]} K 
			- 3 \fIJK (\d_Q C\du{\[\m} J) F\du{\n\r\]} K ~~, ~~~~~   
&(2.4\rma) \cr 
\d_Q H\du{\m\n} I = & - 2 (\Bar\e\g_{\[\m} D_{\n\]} \r^I) 
				+ g (\Bar\e\g_{\m\n} \chi^I ) 
			+ 2 \fIJK D_{\[\m |}\left[ (\d_Q A\du{| \n\]} J ) \varphi^K \right] ~~, 
&(2.4\rmb) \cr 
\d_Q F\du{\m\n} I = & - 2 (\Bar\e\g_{\[\m} D_{\n\]} \l^I) ~~,  
&(2.4\rmc) \cr } $$ 
reflecting the presence of CS terms.   

\doit0{
In ordinary supergravity theories \sg, lagrangians are 
fixed up to quartic fermion terms.  There are two main reasons for this.  First, 
Fierz arrangements to fix quartic fermion terms are too involved.  
Second, due to the $\,\hbox{(fermions)}^2\-$terms in $~\d_Q\hbox{(fermions)}$, even the Paul-terms $~\hbox{(bosons)}\hbox{(fermions)}^2$~ 
contribute to the $\,\e \, \hbox{(boson}) \, \hbox{(fermions)}^2\-$terms in 
$~\d_Q \Lag$, so that computations become too involved.    
For these reasons, we fix our lagrangian up to fermionic quartic terms, just as in 
supergravity \sg.  
} 

\doit0{
Due to the $~\order g1$~ term in (3.6b), even quadratic terms $~\order\phi2$~ 
are non-trivial.  
These terms are either $~\order g 0$~ terms with four sectors: (i) $\l D F$, ~
(ii) $\chi D G$, ~(iii),  $\chi D^2 \varphi$ ~ and (iv) $\r D H$, 
while $~\order g 1$~ terms with three sectors (i) $g \r \, G$, ~(ii) $g \chi H$~ and 
(iii) $g \r D\varphi$.  
At $~\order \phi 3$, we have $~(\hbox{Bosons})^2 \, (\hbox{Fermions})\-$terms, 
which are of $~\order g 0$~ and $~\order g 1$~ types.  
The former set is composed of the 
seven sectors: (i) $~\l G H$,~ (ii) $\chi F H$, ~(iii) $ \r F G$, ~ (iv) $\l H D\varphi$, 
~ (v) $ \r F D\varphi$, ~(vi) $ \l (D H) \, \varphi $, ~(vii) $ \r (D F) \, \varphi$, 
while the latter set is composed of the three sectors:  (i) $ g \l \varphi (D\varphi)$, 
~ (ii) $ g \chi F G$, ~(iii) $g \l G \varphi$.  The only subtlety is the involvement 
of the BIds (3.5) and the non-trivial transformations (3.6).  These sectors are 
also coming from the $~\order\phi2\-$terms, {\it via} the BIds (3.5) after partial 
integrations.  
}

Note that our YMVM and CVM has {\it on-shell} DOF 2+2, while {\it off-shell} 
DOF 3+4, because we have {\it not} added the $~D\-$auxiliary field.  On the other 
hand, our TM is in the {\it off-shell} formulation, because the total off-shell 
DOF is $~4+4$, because the off-shell DOF of 
each field are $~[(4-1)\cdot(4-2)] / 2 = 3$~ for $~B_{\m\n}$, 
$~4$~ for $~\chi$~ and $~1$~ for $~\varphi$.  

The field equations for $~\l^I, ~\chi^I,~\r^I, ~A\du\m I, ~B\du{\m\n} I, ~
\varphi^I$~ and $~C\du\m I$~ are respectively\footnotes{These equations 
are fixed up to $~\order\phi3\,$-terms, due to the quartic fermion terms 
in the lagrangian.}   
$$\li{ & + \Dsl \l^I - \frac 12 g \fIJK \chi^J \varphi^K 
		+ \frac 12 \fIJK (\g^\m \r^J) D_\m \varphi^K \cr 	
& ~~~~~ ~~~~~ ~~~~~ ~~~~~ ~~~~~ ~~~~~ 
		- \frac 14\fIJK (\g^{\m\n} \chi^J) H \du{\m\n} K 
		+  \frac 1{12}\fIJK (\g^{\m\n\r} \r^J) G\du{\m\n\r} K
		 \eqdot 0 ~~, ~~~~~ ~~~ 
&(2.5\rma) \cr 
\noalign{\vskip 0.2in}  
& + \Dsl \chi^I - g \r^I + \frac 12 g \fIJK \l^H \varphi^K 
			- \frac 14 \fIJK (\g^{\m\n} \l^J) H\du{\m\n} K  
			 + \frac 14 \fIJK (\g^{\m\n} \r^J) F\du{\m\n} K \eqdot 0 
			~~, ~~~~~ ~~~
&(2.5\rmb) \cr 
\noalign{\vskip 0.2in}  
& + \Dsl \r^I - g \chi^I + \frac 12 \fIJK (\g^\m\l^J) D_\m\varphi^K \cr 
& ~~~~~ ~~~~~ ~~~~~ ~~ 
		- \frac 1{12} \fIJK (\g^{\m\n\r} \l^J) G\du{\m\n\r} K 
		+ \frac 14 \fIJK (\g^{\m\n} \chi^J) F\du{\m\n} K \eqdot 0
		{~~ , ~~~~~ ~~~~~}
&(2.5\rmc) \cr 
\noalign{\vskip 0.2in}  
& + D_\n F\du\m{\n\, I} + g \fIJK \varphi^J D_\m\varphi^K 
			+ \frac 12 g \fIJK (\Bar\l{}^J \g_\m \l^K) 
			+ \fIJK H\du{\m\n} J D^\n\varphi^K  \cr 
& ~~~~~ ~~~~~ ~~~~~ - \frac 12 \fIJK G\du{\m\r\s} J H^{\r\s\, K}
			+ \frac 12 \fIJK (\Bar\chi{}^J D_\m\r^K) 
		+ \frac 12 \fIJK (\Bar\r{}^J D_\m\chi^K) \eqdot 0 ~~, 		
&(2.5\rmd) \cr 
\noalign{\vskip 0.2in}  
& + D_\r G^{\m\n\r\, I} - g H^{\m\n\, I}  
				- \frac12 \fIJK D_\r (\Bar\l{}^J \g^{\m\n\r} \r^K) \cr 
& ~~~~~ ~~~~~ ~~~~~ ~~~~~ ~~~~~ ~~~~~  
		+ g \fIJK F^{\m\n\, J}  \varphi^K 
		- \frac 12 g \fIJK (\Bar\l{}^J \g^{\m\n} \chi^K) \eqdot 0 ~~, 
&(2.5\rme) \cr 
\noalign{\vskip 0.2in}  
& + D_\m^2 \varphi^I - g \fIJK(\Bar\l{}^J\chi^K) 
	- g^2 \varphi^I - \frac 12 \fIJK F\du{\m\n} J H^{\m\n\, K} \eqdot 0~~, 
&(2.5\rmf) \cr 
\noalign{\vskip 0.2in}  
& + D_\n H^{\m\n\, I} - \frac 12 \fIJK F\du{\r\s} J G^{\m\r\s\, K} 
		- \frac 12 \fIJK (\Bar\chi{}^J D^\m \l^K) 	
		- \frac 12 \fIJK (\Bar\l{}^J D^\m \chi^K) \cr 
& ~~~~~ ~~~~~ ~~~~~ ~~~~~ ~~~~~  ~~~~~ ~~~~~  ~~~~~ ~
		+ \frac 12 g \fIJK (\Bar\l{}^J \g^\m\r^K) 
		- \fIJK F^{\m\n\, J} D_\n\varphi^K \eqdot 0 ~~.   		
&(2.5\rmg) \cr } $$ 
In the derivation of these field equations, we have also used 
other field equations, in order to simply their final expressions, as 
a conventional prescription.  

In the above computation, we do not attempt to fix the $~\order\phi3\-$terms 
in field equations, or equivalently the fermionic $~\order\phi 4 \-$terms in the lagrangian.  There are several remarks about these terms.    
First, our system is non-renormalizable as supergravity theory \sg, as has been 
mentioned after eq.$\,$(2.2).  
Accordingly, the $\,(\hbox{fermion})^2\-$terms in the fermionic transformations such as (2.2b), (2.2e) and (2.2g) are accompanied by the implicit constant  
~$\ell$~ carrying the dimension of $~(\hbox{legnth})$.  In supergravity theory \sg, 
this is the gravitational coupling $~\k$.  In our lagrangian, all 
the quartic-fermion terms carry $~\ell^2$, so that the lagrangian has 
the mass dimension $+4$.  Accordingly, a  
typical Noether-term has the structure $~\ell \,\Psi^2  \,
\partial  \,\Phi$, that produces the terms of the form 
$~\ell^2 \, \e\,\Psi^3 \,\partial  \,\Phi$~ {\it via} $~\d_Q  \,\Psi
\approx  \,\ell \, \e \, \Psi^2$.  Here $~\Psi$~ (or $~\Phi$) is a general fermionic
(or bosonic) fundamental field. These $~\ell^2 \,\e\,\Psi^3 \,\partial  \,\Phi\-$terms are cancelled by the variation of the fermionic quartic terms $~\ell^2\, \Psi^4$, 
{\it via} $~\d_Q \Psi \approx \e \, \partial \Phi $.  
In other words, the structure of these cancellations associated with 
quartic-fermion terms is parallel to supergravity \sg, since $~\ell$~ is analogous 
to $~\k$. 

However, in our peculiar system, this cancellation mechanism may be 
{\it not} simply parallel to conventional supergravity \sg.  
For example, there may be $~\ell^2\Psi^2 \Phi 
\partial\Psi\-$type terms in the action, while $~\ell^2 \e \Psi^2\Phi\-$type terms 
in the transformation rules may exist, because both of
them yield $~\ell^2\e \Psi^3 \partial\Phi\-$type terms, 
canceling each other in $~\d_Q I$.    
At the present time, we do not know, if such terms arise, because the $~\ell^2 \e \Psi^2\Phi\-$type terms in transformations are at $~\order\phi3$, while 
$~\ell^2\Psi^2 \Phi \partial\Psi\-$type terms in the action are at $~\order\phi4$.  
In fact, even in the superspace re-confirmation in the next section, 
we have fixed only the $~\order\phi1$~ and $~\order\phi 2\-$terms in the transformation rules for fermions, such as (3.2d), (3.2e) and (3.2f), 
but {\it not} cubic terms $~\order\phi3$.  Our consistent principle in 
this paper is to fix only $~\order\phi1, ~\order\phi2$~and $~\order\phi3\-$terms in the lagrangian, $~\order\phi1$~ and $~\order\phi2\-$terms in all 
transformation rules, 
while $~\order\phi1$~ and $~\order\phi2\-$terms in all field 
equations.  However, we try to fix {\it neither} $~\order\phi4\-$terms in the lagrangian, {\it nor} $~\order\phi3\-$term in all transformation rules, {\it nor}   
$~\order\phi3\-$terms in all field equations.  
We do {\it not} specify each field meant by $~\phi$~ 
is fermionic or bosonic in this paper, either.   

Second, as an additional difference from supergravity \sg, 
the fermionic quartic terms 
do {\it not} contain any gravitino.  This implies that we can not use the 
conventional technique of `supercovariantizing' 
fermionic field equations.  Due to this feature, as well as the above-mentioned possible non-purely-fermionnic $~\ell^2\Psi^2 \Phi \partial\Psi\-$type terms, 
the quartic terms $~\order\phi4$~ at 
$~\order\ell 2$~ will be more involved than conventional supergravity \sg\ 
which are tedious.
For these reasons, we do not attempt to fix them in this paper. 

Third, according to the past experience in supergravity theory \sg, 
it is understood that the series in terms of $~\k$~ in a lagrangian will stop at a  
finite order, such as the quartic-fermion terms at $~\order\k 2$ \sg.  
However, at the present time, we do {\it not} know, whether this is also the 
case with our globally supersymmetric system.  This is because of the 
above-mentioned differences of our system from supergravity \sg, and 
therefore the analogy with supergravity might be not valid in our system.    
Fourth, since we have already fixed the 
cubic terms in the lagrangian, they seem sufficient for 
non-trivial and consistent couplings as a supersymmetric system.

\bigskip\bigskip 



\leftline{\bf 3.~~Superspace Reformulation of $\,$N=1$\,$ TM} 
\nobreak 

As a reconfirmation of the total consistency of our system, we re-formulate 
our theory in terms of superspace language.  
Our basic superspace BIds for the superfield strengths $~F\du{A B} I, 
~G\du{A B C} I$~ and $~H\du{A B} I$~ are\footnotes{Only in this superspace section, we use the indices $\,{\scst A~=~(a,\a),~B~=~(b,\b),~\cdots}\,$  
for superspace coordinates, where $\,{\scst a,~b,~\cdots~=~0,\,1,\,2,\,3}\,$  
(or $~{\scst \a,~\b,~\cdots~=~1,\,2,\,3,\,4}$) are for 
bosonic (or fermionic) coordinates. 
In superspace, the (anti)symmetrization convention, {\it e.g.,} $X_{\[ A B)} 
\equiv X_{A B} -(-1)^{A B} X_{B A}$~ is different from our 
component notation.}   
$$ \li{ + \fracm16 \nabla_{\[A} G\du{B C D)} I 
		- \fracm 1 4 T\du{\[A B|} E G_{E| C D)} 
			-\fracm 14 f^{I J K} F\du{\[A B} J H\du{C D)} K & \equiv 0 ~~, 
& (3.1\rma) \cr  
+ \fracm 12 \nabla_{\[ A} H\du{ B C)} I 
		- \fracm 12 T\du{\[ A B| } D H\du{ D| C)} I  - g \, G\du{A B C} I 
& \equiv 0  ~~, 
& (3.1\rmb) \cr  
+ \fracm 12 \nabla_{\[ A} F\du{ B C)} I  
		- \fracm 12 T\du{\[ A B| } D F\du{ D| C)} I & \equiv 0 ~~. 
& (3.1\rmb) \cr  } $$ 
These BIds are the superspace generalizations of the component BIds 
(2.3), with the supertorsion terms added for local Lorentz indices, as usual 
in superspace. 

Our basic superspace constraints at mass dimensions $~0 \le d \le 1$~ are 
$$\li{ T\du{\a\b} c = & + 2 (\g^c)_{\a\b}~~, ~~~~
		G\du{\a\b c} I = + 2 (\g_c)_{\a\b} \, \varphi^I~~, 
&(3.2\rma) \cr 
G\du{\a b c}I = & - (\g_{b c}\chi^I )_\a~~, ~~~~
		H\du{\a b} I = - (\g_b\r^I)_\a - f^{I J K} (\g_b\l^J)_\a \, \varphi^K  ~~, 
&(3.2\rmb) \cr 
F\du{\a b} I = & - (\g_b \l^I)_\a ~~, ~~~~ 
		\nabla_\a \varphi^I = - \chi\du\a I ~~,  
&(3.2\rmc) \cr  
\nabla_\a \chi\du\b I = & - \fracm 1 6 (\g^{c d e})_{\a\b} G\du{c d e} I
					- (\g^c)_{\a\b} \nabla_c \varphi^I \cr 
& - \frac 12 f^{I J K} \Big[ + C_{\a\b} (\Bar\l{}^J \r^K) 
				       - (\g_5\g^c)_{\a\b} (\Bar\l{}^J \g_5\g_c\r^K) 
				       - (\g_5)_{\a\b} (\Bar\l{}^J \g_5 \r^K) \Big] 
				       ~~, ~~~~~ 
&(3.2\rmd) \cr  
\nabla_\a \r\du\b I = & + \frac 12 (\g^{c d})_{\a\b} H\du{c d} I 
				   + g \, C_{\a\b} \, \varphi^I 
				   - \frac 12 f^{I J K} (\g^{c d})_{\a\b} F\du{c d} J \varphi^K \cr 
& - \frac 14 f^{I J K} \Big[ + C_{\a\b} (\Bar\l{}^J \chi^K) 
					+ (\g^c)_{\a\b} \, (\Bar\l{}^J \g_c \chi^K) 
					- \frac 12 (\g^{c d})_{\a\b} (\Bar\l{}^J \g_{c d}\chi^K) \cr 
& ~~~~~ ~~~~~ ~~ \,\, - (\g_5\g^c)_{\a\b} (\Bar\l{}^J \g_5\g_c\chi^K) 
					- (\g_5)_{\a\b} (\Bar\l{}^J \g_5 \chi^K) ~~,  
&(3.2\rme) \cr 
\nabla_\a \l\du\b I = & + \frac 12(\g^{c d})_{\a\b} F\du{c d} I 
				  - \frac 12 (\g_5)_{\a\b} \, f^{I J K} (\Bar\r{}^J\g_5\chi^K) ~~. 
&(3.2\rmf) \cr } $$ 
All other components, such as $~G_{\a\b\g}{}^I, ~T\du{\a\b}\g, ~T\du{a b} c, ~
H\du{\a\b} I $~ {\it etc.}~at $~d\le 1$~ are zero.  
Note that $~\hbox{(fermion)}^2\-$terms in (3.2d) through (3.2f) have been determined in superspace by satisfying BIds at ~$d=1$.  
Note that these results are valid up to $~\order\phi3\-$terms, 
which we do not attempt to fix these terms in this paper.  However, all the 
$~\order\phi2\-$terms have been included, as has been also mentioned at the 
end of last section.  

There are also useful relationships obtained from $~d = + 3/2$~ BIds: 
$$\li{ \nabla_\a G_{b c d} = & - \frac12 (\g_{\[ b c} \nabla_{d\]} \chi^I )_\a 
					- \frac 12 f^{I J K} (\g_{\[b | } \l^J)_\a H\du{| c d\]} K 
					+  \frac 12 f^{I J K} (\g_{\[b | } \r^J)_\a F\du{| c d\]} K 
					~~, ~~~~~ ~~~ 
&(3.3\rma) \cr 
\nabla_\a H_{b c}{}^I = & + (\g_{\[ b} \nabla_{c\]} \r^I)_\a 
					- g (\g_{b c} \chi^I)_\a
			 - f^{I J K} \nabla_{\[b} \Big[ (\g_{c\]} \l^J)_\a \varphi^K\Big] ~~, 
&(3.3\rmb) \cr 
\nabla_\a F_{b c}{}^I = & + (\g_{\[ b} \nabla_{c\]} \l^I)_\a ~~,    
&(3.3\rmc) \cr } $$ 
up to $~\order\phi3\-$terms.  
Note the existence of the $~\order\phi 2\-$terms in (3.3a) and (3.3b), 
reflecting the corresponding terms in the component results (2.4a) 
and (2.4b).  

As usual, the satisfaction of all the BIds in superspace by the constraints 
(3.2) and (3.3) is straightforward to perform, from the dimension $~d=0$~ 
to $~d=3/2$, as usual.  In particular, the $\,(\hbox{Fermions})^2\-$terms in 
(3.2d) through (3.2f) are the results of our superspace re-formulation.  

The fermionic $\,\l$~ and $\,\r\-$field equations (2.5a) and (2.5c) are obtained
as usual by computing $~\{ \nabla_\a, \nabla_\b\} \, \l^{\b I}$~ and 
$~\{ \nabla_\a, \nabla_\b\} \, \r^{\b I}$, while the $~\chi\-$field equation is 
shown to be consistent with the component lagrangian.  As has been 
mentioned, since the TM is {\it off-shell} multiplet, we can {\it not} get the 
$\,\chi\-$field equation (2.5b) in superspace directly, but we can show that 
(2.5b) is consistent in superspace.  The bosonic 
field equations (2.5d) - (2.5g) are obtained by applying another fermionic 
derivative on the fermionic field equations (2.5a) - (2.5c).

\bigskip\bigskip 


\leftline{\bf 4.~~Generalization to Non-Adjoint Representations of 
$\,$G = SO(N)} 
\nobreak 

We have so far considered the case for the TM and CVM both carrying only 
the adjoint representation.  We can generalize this result to other more 
general representations, such as an arbitrary real representation of a 
$~SO(N)\-$type gauge group.\footnotes{We can also consider the 
complex representation for $~SU(N)\, $-type gauge groups.}   

To be more specific, we consider the TM $~(B\du{\m\n} i, \chi^i, \varphi^i)$~ 
and the CVM $~(C\du\m i, \r^i)$, where the index $~{\scst i}$~ is for any 
real representation of a gauge group $~G = SO(N)$.  Let $~(T^I)^{j k}$~ 
be the generator of the group $~G$.  Then    
our action $~I' \equiv \int d^4 x \, \Lag'$~ has the lagrangian\footnotes{Since 
the metric for the gauge group $~G = SO(N)$~ is positive definite, 
we do {\it not} distinguish the upper or lower indices for $~{\scst i,~j,~\cdots~=~
1,~2,~\cdots ,~\hbox{dim} \, R}$, where $~{\bf R}$~ is 
a real representation of $~G$.}  
$$\li{ \Lag ' = & - \frac 1{12} ( G\du{\m\n\r} i)^2
			 + \frac 12 (\Bar\chi{}^i \Dsl\chi^i) 
			- \frac 12 (D_\m\varphi^i)^2  
	- \frac 12 g^2 ( \varphi^i)^2 - g (\Bar\r^i \chi^i) \cr 
& - \frac 14 (H\du{\m\n} i)^2 + \frac 12 (\Bar\r{}^i \Dsl\r^i) 
   	 - \frac 14 (F\du{\m\n} I)^2 + \frac 12 (\Bar\l{}^I \Dsl\l^I) \cr  
& - \frac12 g (T^I)^{j k} (\Bar\l{}^I \chi^j) \, \varphi^k  
		+ \frac12 (T^I)^{j k} (\Bar\l^I \g^\m \r^j) D_\m \varphi^k 
		+ \frac1{12} (T^I)^{j k} (\Bar\l^I \g^{\m\n\r} \r^j ) \, G\du{\m\n\r} k  
		~~~~~ ~~\cr  
& + \frac 14 (T^I)^{j k}(\Bar\r{}^j \g^{\m\n} \chi^k) F\du{\m\n} I 
		- \frac 14 (T^I)^{j k} (\Bar\l^I \g^{\m\n} \chi^j) H_{\m\n}{}^k 
		- \frac 12 (T^I)^{j k} F\du{\m\n} I H^{\m\n\, j} \varphi^k ~~, 
& (4.1) \cr} $$  
up to quartic terms $~\order\phi4$.  
Our action $~I'$~ is invariant under global $~N=1$~ supersymmetry 
$$\li{ \d_Q B\du{\m\n} i = & + (\Bar\e\g_{\m\n} \chi^i) 
			- 2 (T^J)^{i k} C\du{\[\m |} k (\d_Q A\du{| \n\]} J )  ~~, 
&(4.2\rma) \cr  
\d_Q \chi^i = & + \frac 16 (\g^{\m\n\r} \e) G_{\m\n\r}{}^i  
				- (\g^\m\e) D_\m \varphi^i \cr 
& - \frac 12 (T^J)^{i k} \Big[ + \e (\Bar\l{}^J \chi^k) 
					- (\g_5\g^\m \e) (\Bar\l{}^J \g_5 \g_\m \chi^k) 
					  -  (\g_5 \e) (\Bar\l{}^J \g_5\chi^k)\Big] ~~, 
&(4.2\rmb) \cr 
\d_Q \varphi^i = & + (\Bar\e\chi^i) ~~, 
&(4.2\rmc) \cr  
\d_Q C\du\m i =  & + (\Bar\e\g_\m\r^i) 
		- (T^J)^{i k} (\Bar\e\g_\m \l^J) \varphi^k ~~, 
&(4.2\rmd) \cr  
\d_Q \r^i =  & + \frac 12 (\g^{\m\n}\e) H\du{\m\n} i - g \e \varphi^i  
		+  \frac 12 (T^J)^{i k} (\g^{\m\n}\e) F\du{\m\n} J \varphi^k \cr 
& - \frac 14 (T^J)^{i k} \Big[ + \e (\Bar\l{}^J \chi^k) 
			- (\g^\m\e) (\Bar\l{}^J \g_\m\chi^k) 
			+ \frac 12 (\g^{\m\n}\e) (\Bar\l{}^J \g_{\m\n} \chi^k )  \cr 
& ~~~~~ ~~~~~ ~~~~~  - (\g_5\g^\m\e) (\Bar\l{}^J \g_5 \g_\m\chi^k ) 
			- (\g_5 \e) (\Bar\l{}^J \g_5\chi^k ) \Big] ~~, 
&(4.2\rme) \cr
\d_Q A\du\m I =  & + (\Bar\e\g_\m\l^I) ~~, 
&(4.2\rmf) \cr 
\d_Q \l^I = & + \frac12 (\g^{\m\n}\e) F\du{\m\n} I 
			- \frac 12 (T^I)^{j k} (\g_5\e) (\Bar\r^j\g_5 \chi^k) ~~.    
&(4.2\rmg) \cr } $$  

The essential point is that all the cubic-order terms 
contain one component field $~A\du\m I$~ or $~\l^I$~ with the index $~{\scst I}$, 
and the remaining two component fields out of either TM or CVM 
carry the indices $~{\scst j}$~ and $~{\scst k}$.  
So the cancellation structure is parallel to the
adjoint-representation case, {\it e.g.,} 
with the structure constant $~\fIJK$~ replaced by the matrix $~- (T^J)^{i k}$~ in    
$~D_\m \chi^I = \partial_\m\chi^I + g \fIJK A\du\m J \chi^K ~~\Longrightarrow~~ 
D_\m \chi^i = \partial_\m\chi^i - g (T^J)^{i k} A\du\m J \chi^k$.  
Accordingly, the Stueckelberg mechanism \stueckelberg\ works in a 
parallel fashion, because $~C\du\m i$~ is absorbed into the longitudinal 
component of $~B\du{\m\n} i$, both in the same representation $~{\bf R}$.

\bigskip\bigskip 


\leftline{\bf 5.~~Coupling to $~N=1$~ Supergravity} 
\nobreak 

Once we have established the $~N=1$~ global system of non-Abelian 
TM with non-trivial and consistent interactions, 
the next natural step is to make $~N=1$~ supersymmetry {\it local},  
coupling to $~N=1$~ supergravity.  

This coupling is rather straightforward, because most of the basic structure is 
parallel to the usual matter coupling to supergravity, 
except for certain couplings to be mentioned later.  
Our result for the lagrangian $~\Tilde\Lag$~ of our action is 
$~ \Tilde I \equiv \int d^4 x\, \Tilde \Lag$: 
$$ \li{ e^{-1} \Tilde \Lag = & - \frac 14 R (\o) -  \left[ \Bar\psi_\m\g^{\m\n\r} 
			D_\n(\o) \psi_\r \right]  
		- \frac 1{12} ( G\du{\m\n\r} I)^2 
			+ \frac 12 \big[ \, \Bar\chi{}^I \Dsl(\o)\chi^I \big]  
				- \fracm12 (D_\m\varphi^I)^2 \cr 
& - \frac 14 (F\du{\m\n} I)^2 + \frac 12 \big[ \, \Bar\l{}^I \Dsl \l^I \big]  
			- \frac 14 (H\du{\m\n}I)^2 
			+ \frac12 \big[ \, \Bar\r{}^I \Dsl (\o) \r^I \big] 
			- g (\Bar\chi{}^I \r^I) - \frac 12 g^2 (\varphi^I)^2 \cr 
& - \frac 12 g f^{I J K} (\Bar\l{}^I \chi^J) \varphi^K 
		- \frac 14 f^{I J K} (\Bar\l{}^I \g^{\m\n} \chi^J) H\du{\m\n} K \cr 
& + \frac 1{12} f^{I J K} (\Bar\l {}^I \g^{\m\n\r} \r^J) G\du{\m\n\r} K 
		+ \frac 14 f^{I J K} (\Bar\r{}^I \g^\m \chi^J) F\du{\m\n} K \cr 
& - \frac 12 f^{I J K} F\du{\m\n} I H^{\m\n\, J} \varphi^K 
			+ \frac 12 f^{I J K} (\Bar\l{}^I \g^{\m\n} \r^J) D_\m \varphi^K \cr 
& + (\Bar\psi_\m\g^\n\g^\m \chi^I) D_\n \varphi^I 
		+ \frac 16 (\Bar\psi_\m \g^{\r\s\t} \g^\m \chi^I) G\du{\r\s\t} I \cr 
& - \frac 12 (\Bar\psi_\m \g^{\r\s} \g^\m \l^I) F\du{\r\s} I 
		- \frac 12 (\Bar\psi_\m \g^{\r\s}\g^\m\r^I ) H\du{\r\s} I 
		- g (\Bar\psi_\m\g^\m \r^I) \varphi^I ~~, 
&(5.1) \cr } $$ 
up to $~\order\phi4$~ terms.  		

Our action $~\Tilde I$~ is now invariant under local $~N=1$~ 
supersymmetry
$$\li{ \d_Q e \du\m m = & - 2 (\Bar\e\g^m\psi_\m) ~~, 
&(5.2\rma) \cr  
\d_Q \psi_\m = & + D_\m(\Hat\o) \e  
			- \frac16 (\g\du\m{\r\s\t} \e) \Hat G\du{\r\s\t} I \varphi^I ~~,  
&(5.2\rmb) \cr  
\d_Q B\du{\m\n} I = &  + (\Bar\e\g_{\m\n} \chi^I) 
					- 2 \fIJK C\du{\[\m | } J (\d_Q A\du{| \n\]} K) 
					- 4 (\Bar\e \g_{\[\m} \psi_{\n\]}) \varphi^I ~~, 
&(5.2\rmc) \cr  
\d_Q \chi^I = & + \frac 16 (\g^{\m\n\r} \e) \Hat G_{\m\n\r}{}^I 
				- (\g^\m\e) \Hat D_\m \varphi^I \cr 
& + \frac 12 \fIJK \Big[ + \e (\Bar\l{}^J\r^K) 
					- (\g_5\g^\m \e) (\Bar\l{}^J \g_5\g_\m \r^K) 
					  -  (\g_5 \e) (\Bar\l{}^J \g_5\r^K)\Big] ~~, 
&(5.2\rmd) \cr  
\d_Q \varphi^I = & + (\Bar\e\chi^I) ~~, 
&(5.2\rme) \cr  
\d_Q C\du\m I =  & +(\Bar\e\g_\m\r^I) 
		+ f^{I J K} (\Bar\e \g_\m \l^J) \varphi^K ~~, 
&(5.2\rmf) \cr  
\d_Q \r^I =  & + \frac 12 (\g^{\m\n}\e) \Hat H\du{\m\n} I - g \, \e \, \varphi^I  
		- \frac 12 f^{I J K} (\g^{\m\n}\e) \Hat F\du{\m\n} J\varphi^K \cr 
& + \frac 14 f^{I J K} \Big[ + \e (\Bar\l{}^J \chi^K) 
			- (\g^\m\e) (\Bar\l{}^J \g_\m\chi^K ) 
			+ \frac 12 (\g^{\m\n}\e) (\Bar\l{}^J \g_{\m\n} \chi^K )  \cr 
& ~~~~~ ~~~~~ ~~\,\, - (\g_5\g^\m\e) (\Bar\l{}^J \g_5 \g_\m\chi^K ) 
			- (\g_5 \e) (\Bar\l{}^J \g_5\chi^K ) \Big] ~~, 
&(5.2\rmg) \cr
\d_Q A\du\m I =  & + (\Bar\e\g_\m\l^I) ~~, 
&(5.2\rmh) \cr 
\d_Q \l^I = & + \frac12 (\g^{\m\n}\e) \Hat F\du{\m\n} I 
			+ \frac 12 \fIJK (\g_5\e) (\Bar\r^J\g_5 \chi^K) ~~,     
&(5.2\rmi) \cr } $$  
up to $~\order\phi 3$~ terms. 
The supercovariant field strengths are defined as usual in supergravity 
\sg\ by 
$$ \li{ \Hat F\du{\m\n} I \equiv & + 2 \partial_{\[\m} A\du{\n\]} I 
			+ g \fIJK A\du\m J A\du\n K 
					- 2(\Bar\psi_{\[\m} \g_{\n\]} \l^I) 
		= F\du{\m\n} I - 2(\Bar\psi_{\[\m} \g_{\n\]} \l^I) ~~, 
&(5.3\rma) \cr 
\Hat G\du{\m\n\r} I \equiv & + 3 D_{\[\m} B\du{\n\r\] } I 
				- 3 \fIJK C\du{\[\m} J F\du{\n\r\]} K 
					-3 (\Bar\psi_{\[\m} \g_{\n\r\]} \chi^I)  
					+ 6 (\Bar\psi_{\[\m|} \g_{|\n|} \psi_{|\r\]} ) \varphi^I \cr 
= & + G\du{\m\n\r} I -3 (\Bar\psi_{\[\m} \g_{\n\r\]} \chi^I ) 
					+ 6 (\Bar\psi_{\[\m|} \g_{|\n|} \psi_{|\r\]} ) \varphi^I ~~, 
&(5.3\rmb) \cr 
\Hat H\du{\m\n} I \equiv & + 2 D_{\[\m} C\du{\n\]} I + g B\du{\m\n}I 
		- 2 (\Bar\psi_{\[\m} \g_{\n\]} \r^I) 
		= H\du{\m\n} I - 2 (\Bar\psi_{\[\m} \g_{\n\]} \r^I) ~~, 
&(5.3\rmc) \cr 
\Hat D_\m\varphi^I \equiv & + D_\m\varphi^I - (\Bar\psi_\m\chi^I) ~~.
&(5.3\rmd) \cr } $$ 

Certain remarks are in order.  First, the  
last term in (5.1) of the type $~g (\Bar\psi \g\r) \varphi$~ is related to 
the $~\varphi\-$linear term in $~\d_Q \r$~ in (5.2g).  Second, the
$~\d_Q B_{\m\n}$~ contains the $~( \Bar\e \g\psi) \varphi\-$term.   
This is consistent with $~G\du{\a\b c} I = + 2 (\g_c)_{\a\b} \, \varphi^I$~ 
in (3.2a) in superspace.  Third, for the $~g \psi\r\chi\-$terms, we need 
non-trivial Fierz rearrangement.  To be more specific, 
there are three contributions to this sector: 
(i)$~g (\Bar\psi\g\r) \varphi$, (ii)~$g e (\Bar\chi\r)$, and 
(iii) ~$ (\Bar\psi\g\g\r) H\-$terms.  This rearrangement is highly non-trivial,   
showing the consistency of our total system.  

As the couplings to supergravity in (5.1) show, our original {\it globally} 
supersymmetric system shares certain feature with supergravity, 
such as fermionic bilinear terms.  Because such terms are 
common in supergravity \sg, but {\it not} in conventional global supersymmetry.  
Our original {\it global} system already possessed the feature of {\it local} 
$~N=1$~ supersymmetry.  As has been mentioned after (2.2), 
the conventional dimensional analysis tells that such terms imply 
non-renormalizability.  
In other words, our {\it globally} supersymmetric system already had 
a hidden gravitational constant $~\k$~ providing negative mass dimension.     
In a sense, this feature resembles $~\s\-$models with 
non-renormalizable couplings, sharing certain features  
with gravity interactions.

\bigskip\bigskip 



\leftline{\bf 6.~~Concluding Remarks} 
\nobreak 

In this paper, we have carried out the $~N=1$~ 
supersymmetrization in 4D of a non-Abelian tensor with 
consistent couplings, as a special case \chu\ of the minimal tensor 
hierarchy discussed in \ssw, which is further a special case of more general 
hierarchy in \dws\dwns. 
We have given both the component and superspace 
formulations of our system, providing the non-trivial consistency of our 
system.  Our CVM $~(C\du\m I, \r^I)$~ plays the role of a Stueckelberg 
\stueckelberg\ compensator multiplet, being absorbed into the 
TM $~(B\du{\m\n} I, \chi^I, \varphi^I)$, making the latter massive. 

We have also generalized the adjoint-representation case to the 
general real representation for $~ G = SO(N)$.  The action invariance 
works in a fashion parallel to the former.  We foresee no obstruction against 
generalizing these result further to the complex representation of, {\it e.g.,} 
$~G= SU(N)$~ group.  
Finally, we have also coupled the global $~N=1$~ system to $~N=1$~ 
supergravity up to quartic terms.  
This has provided a non-trivial confirmation for the 
total consistency of the non-Abelian TM.  

\doit0{
Our formulation has solved problems in supersymmetric 
gauge field theories, and has given a new system, 
based on a very simple field content.  First, 
we have established the supersymmetric generalization of the non-Abelian tensor 
$~B\du{\m\n} I$~ with consistent couplings in explicit lagrangians.  
Second, we have solved the 
common problem with a vector field $~C\du\m I$~ carrying an adjoint index, 
which is {\it not} the gauge field of the gauge group $~G$~ itself.  
The solution turned out to be the introduction of an extra vector $~C\du\m I$~ playing a role of Stueckelberg compensator, eventually absorbed into the longitudinal components of the non-Abelian tensor 
$~B\du{\m\n} I$.  In other words, the former is collaborating with 
the latter in a Stueckelberg mechanism \stueckelberg, avoiding the common 
consistency problem of couplings.    
In fact, our coupling constant $~g$~ coincides with  
the mass of the TM.  This implies that the consistent couplings for 
the non-Abelian TM and its mass {\it via} the Stueckelberg 
mechanism \stueckelberg\ 
are closely related to each other.  Third, the adjoint index on 
the non-gauge vector field $~C\du\m I$~ is further generalized to 
an arbitrary real representation index of $~G = SO(N)$.  Fourth, 
most importantly, we have carried out the supersymmetrization of such 
a Stueckelberg mechanism for a non-Abelian tensor.  Fifth, even though 
our {\it algebra} with $~\d_\a, ~\d_\b$~ and $~\d_\g$~ is indeed a special 
case of the hierarchy in \dws, we have given explicit lagrangians with the 
{\it physically propagating vector field} $~C\du\m I$~ that has not been 
presented before. 
}

It has been known that certain problem exists in the 
quantization of Stueckelberg model \stueckelberg\   
for non-Abelian gauge groups 
\ref\problemnonabelian{\problemnonabeliancont}. 
The common problem is that the longitudinal components of the gauge field 
do not decouple from the physical Hilbert space, 
upsetting the renormalizability and unitarity of the system 
\problemnonabelian.  For this issue, we clarify our standpoints as follows:    
First of all, our theory is {\it not} renormalizable from the outset, 
due to Pauli couplings.  Our theory makes stronger 
sense, when couplings to supergravity are also taken into account, as 
we have done in section 5.  
Moreover, there are certain theories in 4D, such as 
non-linear sigma models which are {\it not} renormalizable, but are 
{\it not} excluded from the outset.  
So we do {\it not} go into the renormalizability issue in this paper.  
Second, thanks to $~N=1$~ supersymmetry, 
our system has good chance to have a better quantum behavior, 
compared with non-supersymmetric systems.  

As will be shown in Appendix A, the purely bosonic 
part of our system can be generalized to  
arbitrary space-time dimensions with arbitrary signatures.  
The key ingredient is the tensor 
$~B\du{\m_1\cdots\m_{p+1}} I$~ and a Stueckelberg-type \stueckelberg\ 
compensator $~C\du{\m_1\cdots\m_p}I$.  

The potential importance of the result in this paper is  
$~N=1$~ supersymmetry that has better quantum behavior compared with  
non-supersymmetric cases.   
We have presented a new {\it supersymmetric} physical system with 
Stueckelberg mechanism that solves both 
the problem with non-Abelian tensor, and the problem with 
extra vector fields in the non-singlet representation of a non-Abelian gauge group.




\bigskip\bigskip

\doit0{
We are grateful to W.~Siegel for important discussions, and reading 
the manuscript.}

\doit1{This work is supported in part by Department of Energy 
grant \# DE-FG02-10ER41693.  
} 

\doit0{
We are grateful to the referees of this paper 
for important suggestions to improve our paper.  
Our research is   
} 

\bigskip\bigskip\bigskip




\leftline{\bf Appendix A:~~Higher-Dimensional Application 
of Purely Bosonic System} 
\nobreak 


In this appendix, we generalize the purely bosonic part of our system in 4D 
into arbitrary space-time dimensions with arbitrary signatures.  
We also apply it to the case of tensor-vector duality in 6D, and perform a DR to 4D.  
Our field content is $~(A\du\m I, B\du{\[n-1\]} I, 
C\du{\[n-2\]} I)$.\footnotes{We use the symbols like $~{\scst \[ n\]}$~ 
for totally antisymmetric indices $~{\scst \m_1\m_2\cdots\m_n}$~ 
in order to save space.} 

We generalize the definitions of field strengths (2.1a) and (2.1b)  
to arbitrary space-time dimension $~D$~ as 
$$\li{G\du{\m_1\cdots\m_n} I  
& \equiv + n D_{\[ \m_1} B\du{\m_2\cdots\m_n \]} I 
			- \fracm{n(n-1)} 2 f^{I J K} 
			C\du{\[ \m_1\cdots\m_{n-2}} J F\du{\m_{n-1} \m_n \] } K~~, 
&(\rmA.1\rma) \cr 
H\du{\m_1\cdots\m_{n-1}} I & \equiv 
		+(n-1) D_{\[ \m_1} C\du{\m_2\cdots\m_{n-1} \] } I 
		+ g B\du{\m_1\cdots\m_{n-1}}I ~~.  
&(\rmA.1\rmb) \cr } $$ 
The YM field strength $~F$~ is the same as in (1.2).  
The BIds for these field strengths are 
$$\li{ D_{\[\m} F\du{\n\r \]} I & \equiv 0 ~~, 
&(\rmA.2\rma) \cr 
D_{\[\m_1} G\du{\m_2\cdots\m_{n+1}\]}I 
& \equiv + \fracm n 2 f^{I J K} 
			F\du{\[\m_1\m_2|} J H\du{| \m_3\cdots\m_{n+1}\]} K ~~, 
&(\rmA.2\rmb) \cr 
D_{\[\m_1} H\du{\m_2\cdots\m_n\]}I 
& \equiv + \fracm 1 n \, g\, G\du{\m_1\cdots\m_n} I ~~. 
&(\rmA.2\rmc) \cr } $$ 

The $~\a,~\b$~ and $~\g\-$transformations for  
$~A\du\m I, ~B\du{\[n-1\]} I $~ and $~C\du{\[n-2\]} I$~ are   
the generalizations of our 4D case: 
$$\li{ & \d_\a (A\du\m I , ~B\du{\[n-1\]} I , ~C\du{\[n-2\]} I ) 
		= (D_\m \a^I , ~ - g f^{I J K} \a^J B\du{\[n-1\]} K, 
			~ - g f^{I J K} \a^J C\du{\[n-2\]} K) ~~, ~~~~~ ~~~~~ 
&(\rmA.3\rma) \cr 
& \d_\a (F\du{\m\n} I , ~G\du{\[n\]} I , ~H\du{\[n-1\]} I) 
		=  - g f^{I J K} \a^J (F\du{\m\n} K , ~G\du{\[n\]} K , ~H\du{\[n-1\]} K)~~,   
&(\rmA.3\rmb) \cr 
& \d_\b B\du{\m_1\cdots\m_{n-1}} I 
			= + (n-1) D_{\[\m_1} \b\du{\m_2\cdots\m_{n-1}\]} I ~~, 
			~~~~ \d_\b A\du\m I = 0 ~~, 
&(\rmA.3\rmc) \cr 
& \d_\b C\du{\m_1\cdots\m_{n-2}} I 
			= - g \b\du{\m_1\cdots\m_{n-2}} I ~~,  
&(\rmA.3\rmd) \cr 
& \d_\b (F\du{\m\n} I, ~G\du{\[n-1\]} I, ~H\du{\[n-2\]} I) = 0 ~~,  
&(\rmA.3\rme) \cr
& \d_\g C\du{\m_1\cdots\m_{n-2}} I  
			= +(n-2) D_{\[\m_1} \g\du{\m_2\cdots\m_{n-2}\]} I ~~, 
					~~~~~ \d_\g A\du\m I = 0 ~~, 
&(\rmA.3\rmf) \cr 
& \d_\g B\du{\m_1\cdots\m_{n-1} } I 
			= + \fracm{(n-1)(n-2)} 2 f^{I J K} \, \g\du{\[ \m_1\cdots\m_{n-3} | } J 
		F\du{| \m_{n-2} \, \m_{n-1}\] } K ~~, ~~~~~ 
&(\rmA.3\rmg) \cr 
& \d_\g (F\du{\m\n} I, ~G\du{\[n-1\]} I, ~H\du{\[n-2\]} I) = 0 ~~. 
&(\rmA.3\rmh) \cr } $$ 
Eq.~(\rmA.3d) shows that the $~C\-$field is  
a Stueckelberg field absorbed into the longitudinal components of the 
$\, B\-$field.  

A typical action $~I \equiv \int d^D x\, \Lag $~ is given by the lagrangian  
$$\li{ \Lag = & - \fracm 1{2(n!)} (G\du{\[n\]} I )^2 
				 - \fracm 1{2\cdot(n-1)!} (H\du{\[n-1\]} I )^2
				 - \fracm 14 (F\du{\m\n} I)^2 ~~, 
&(\rmA.4) \cr } $$  
yielding the $~B$~ and $~C\-$field equations 
$$ \li{ \fracmm{\d\Lag}{\d B\du{\[n-1]} I} 
& = \fracm1{(n-1)!} \left( D_\m G^{\m\[n-1\] \, I } 
			- g H^{\[n-1\]\, I} \right) \eqdot ~ 0 ~~, 
&(\rmA.5\rma) \cr 
\fracmm{\d\Lag}{\d C\du{\[n-2\]}I } 
& = \fracmm1{(n-2)!} \left( D_\n H^{\n \[n-2\]\, I} 
			+ \fracm 12 f^{I J K} F\du{\r\s} J G^{\[n-2\] \r\s \, K} \right) 
			\eqdot ~ 0 ~~. 
&(\rmA.5\rmb) \cr } $$  
As in the 4D case, it is straightforward to show the consistency 
$$ \li { 0 \eqques D_\m \left( \fracmm{\d\Lag}{\d B\du{\m \[n-2\]} I} \right) 	
& \equiv - \fracm1{n-1} g \left( \fracmm{\d\Lag}{\d C\du{\[ n-2\] }I} \right) 
			\eqdot ~ 0 ~~, 
&(\rmA.6\rma) \cr 
0 \eqques D_\m \left( \fracmm{\d\Lag}{\d C\du{\m\[n-3\]}I} \right) 	
& \equiv + \fracm{n-1} 2 f^{I J K} F\du{\r\s} J 
		\left( \fracmm{\d\Lag}{\d B\du{\[n-3\] \r\s}K} \right) 
		\eqdot ~ 0  ~~~~\qed~~~~~ 
&(\rmA.6\rmb) \cr } $$ 


We next apply our result to $~6D$~ with the signature $~(-,-,+,+,+,+)$, and  
consider the duality condition 
$$\li{ F\du{\m\n} I & \eqstar + \fracm 1{24}\, \e\du{\m\n}{\r\s\t\l} 
 					\, G\du{\r\s\t\l} I ~~, ~~~~~ 
G\du{\m\n\r\s} I \eqstar + \fracm 1 2 \,\e\du{\m\n\r\s}{\t\l} \, F\du{\t\l} I ~~.  
&(\rmA.7) \cr} $$  
This duality looks similar to eq.~(3.6) in \ssw, but the existence of the 
physical scalar field $~\phi^I$~ in the latter makes the fundamental difference.   

We have to first confirm the consistency of (A.7) with the $~G$~ and $~H\-$BIds.  
First, the rotation of the 2nd equation in (A.7) gives   
$$\li{ 0 & \eqques + \e^{\m\n\r\s\t\l}  
	         D_\n \left( G\du{\r\s\t\l} I  
			 - \fracm 12 \e\du{\r\s\t\l}{\o\psi} F\du{\o\psi} I \right)  
 		\equiv + \e^{\m\n\r\s\t\l} \left( 2 f^{I J K} F\du{\n\r} J H\du{\s\t\l} K \right)
		- 24 D_\n F^{\m\n \, I} \cr
& = - 24 \left( D_\n F^ {\m\n\, I} 
 		 -\fracm1{12} \e^{\m\n\r\s\t\l} 
		 	f^{I J K} F\du{\n\r} J H\du{\s\t\l} K \right) ~~. 
&(\rmA.8) \cr} $$
In the second identity in (A.8), we have used the $~G\-$BId (A.2b).     
The first term in the last line is the kinetic term of $~A\du\m I$, so that its   
last term is its source term.  
Second, in order to see if eq.~(A.8) has consistent solutions, 
we can confirm the conservation of the source term, by 
applying $~D_\m$~ on (A.8) based on $~H\-$BId (A.2c) and (A.7), 
but we skip the details here.   

We next show that the usual self-duality relationship in $~D=2+2$~ 
$$\li{ & F\du{\m\n} I \eqstar + \fracm 12 \,\e\du{\m\n}{\r\s} \, F\du{\r\s} I ~~
&(\rmA.9) \cr} $$  
is embedded into (A.7).  To this end, we use {\it hat} symbols both 
on fields and indices in 6D, while {\it no hats} on 4D quantities 
from now on.  We also use 
$~{\scst \hat\m,~\hat\n,~\cdots~=~1,~2,~3,~4,~5,~6}$~ and $~{\scst\m,~\n,~\cdots~=~1,~2,~3,~4}$, while $~{\scst \a,~\b,~\cdots~=~5,~6}$.  
Our basic ans\" atze for the DR are 
$$\li{ \Hat G\du{\hat\m\hat\n\hat\r\hat\s} I 
	\eqstar & + \Hat F \du{\[ \hat\m\hat\n } I \Hat P_{\hat\r\hat\s \]} ~~, 
	~~~~~ \Hat P_{\hat\m\hat\n} 
\equiv + \Hat\partial_{\hat\m} \Hat X_{\hat\n} 
				- \Hat\partial_{\hat\n} \Hat X_{\hat\m} ~~, 
		~~~~ \Hat H\du{\hat\m\hat\n\hat\r} I 
		\eqstar + \fracm 12 g \Hat F\du{\[\hat\m\hat\n} I \Hat X_{\hat\r \]} 
		~~ , ~~~~~ ~~~~~ ~~ 
&(\rmA.10\rma) \cr 
\Hat P_{\hat\m\hat\n} = & \e_{\a\b} ~~(\hbox{for}~~
								{\scst \hat\m ~= ~\a, ~
										  \hat\n ~= ~\b})~~, ~~~~  
\Hat F\du{\hat\m\hat\n} I = \Hat F\du{\m\n} I = F\du{\m\n} I  
						~~~~ (\hbox{for} ~~ 
						{\scst \hat\m~=~\m, ~\hat \n~=~\n} ) ~~, 
						~~~~~ ~~~~~ 
&(\rmA.10\rmb) \cr 
\Hat\e\,{}^{\hat\m\hat\n\hat\r\hat\s\hat\t\hat\l} 
		= & \Hat\e\,{}^{\m\n\r\s\a\b} = \e^{\m\n\r\s} \e^{\a\b} ~~~~ 
						(\hbox{for}~~ 
						{\scst \[ \hat\m\hat\n\hat\r\hat\s\hat\t\hat\l \] 
						~=~ \[ \m\n\r\s\a\b\] }) ~~.   
&(\rmA.10\rmc) \cr } $$ 
Other components, such as $~\Hat P_{\m\b}$~ are all zero.  
We can confirm that (A.10) are consistent with the BIds (A.2b) and (A.2c).     
It is easy to show that the 
$~{\scst \[ \a\b\]}$~ and $~{\scst \[\m \a\]}\-$components 
of the first equation in (\rmA.7) are satisfied, 
while the $~{\scst \[\m\n\]}\-$component  
gives directly the 4D self-duality (A.9).  
Thus the 4D self-duality $~F \eqstar \Tilde F$~ 
is indeed embedded in the 6D duality (\rmA.7).  

We next generalize the 6D result to the $~D=2m+2$~ 
with the signature $~(-,-,\overbrace{+,\cdots,+}^{2m})$.  
The duality condition (\rmA.7) is generalized to 
$$ \li{ & \Hat F\du{\hat\m\hat\n} I 
\eqstar + \fracm1{(2m)!} \, 
		\Hat\e\du{\hat\m\hat\n}{\hat\r_1\cdots\hat\r_{2m}} 
		\, \Hat G\du{\hat\r_1\cdots\hat\r_{2m}} I ~~, ~~~~
		\Hat G\du{\hat\r_1\cdots\hat\r_{2m}} I 
		\eqstar + \fracm 12 \, \Hat \e\du{\hat\r_1\cdots\hat\r_{2m}} 
			{\hat\m\hat\n}  \, \Hat F\du{\hat\m\hat\n} I ~~. 
&(\rmA.11) \cr } $$ 
As in the 6D case, we can first confirm the consistency with BIds.  
We can next confirm the current conservation, whose details are skipped here.

The previous ans\" atze for 6D case in (A.10) are generalized to 
$$ \li{ \Hat G\du{\hat\m_1\cdots\hat\m_{2m}} I 
			\eqstar & + c \Hat F\du{\[ \hat\m_1 \hat\m_2 | } I 
					\Hat P^{(1)}_{| \hat\m_3\hat\m_4 |} \cdots 
					\Hat P^{(m-1)}_{| \hat\m_{2m-1} \, \hat\m_{2m} \] } ~~,
		~~~~\Hat P^{(k)}_{\hat\m\hat\n} 
				\equiv \Hat\partial_{\hat\m} \Hat X^{(k)}_{\hat\n}
					- \Hat\partial_{\hat\n} \Hat X^{(k)}_{\hat\m}
					~~,~~~~~ ~~~ 
&(\rmA.12\rma) \cr 	
\Hat H\du{\hat\m_1\cdots\hat\m_{2m-1}} I 
			\eqstar & + \fracm 1 m c g \Hat F\du{\[ \hat\m_1\hat\m_2| } I 
					\Hat P^{(1)}_{| \hat\m_3 \hat\m_4 | } \cdots 
					\Hat P^{(m-2)}_{| \hat\m_{2m-3} \, \hat\m_{2m-2} | } 
					\Hat X_{| \hat\m_{2m-1} \]} ~~, 
&(\rmA.12\rmb) \cr 
\Hat P^{(k)}_{\hat\m\hat\n} = &  \Hat P^{(k)}_{2k+3, ~2k+4} 
			= - \Hat P^{(k)}_{2k+4, ~2k+3}
			= \e^{(k)}_{2k+3,~ 2k+4} 
			= - \e^{(k)}_{2k+4,~ 2k+3} = + 1	 ~~ \cr 
&  (\hbox{for}~~
			{\scst \hat\m ~=~ 2k+3,~\hat\n~=~2k+4};~ 
			{\scst k~=~1,~\cdots,~m-1)} ~,  
&(\rmA.12\rmc) \cr 
\Hat F\du{\hat\m\hat\n} I  = & F\du{\m\n} I ~~~~ 
(\hbox{for} ~~{\scst \hat\m~=~\m,~\hat\n~=~\n}) ~~, 
&(\rmA.12\rmd) \cr 
\Hat \e^{\hat\m_1\cdots\hat\m_{2m+2}}  
= & \e^{\m\n\r\s} \, \e^{\a_1\cdots\a_{2m-2}} 
		= \e^{\m\n\r\s} \, \e_{(1)}^{\[ \a_1\a_2 | } \cdots 
		\, \e_{(m-1)}^{| \a_{2m-3} \a_{2m-2} \] } \cr 
&  ( \hbox{for}~~ 
				{\scst \[\hat\m_1 \cdots \hat\m_{2m+2} \]  
				~=~ \[ \m\n\r\s\a_1\cdots\a_{2m-2} \] } )  ~~. 
& (\rmA.12\rme) \cr } $$ 
where $~c$~ is a constant to be fixed later.  

As before, we can also confirm the ~$G$~ and $~H\-$BIds for (A.11).    
The constant $~c$~ in (A.12a) is fixed by getting the 
4D self-duality in the $~{\scst\[\m\n\]}\-$component of the first equation 
in (A.11):    
$$ \li{ F\du{\m\n}I \eqstar & + \fracm1{(2m)!} 
				\, \Hat\e\du{\m\n}{\hat\r_1\cdots\hat\r_{2m}} 
				\, \Hat G\du{\hat\r_1\cdots\hat\r_{2m}} I  
		= + \fracm{{{2m}\choose 2}}{(2m)!} \, 
			\Hat \e\du{\m\n}{\r\s\a_1\cdots\a_{2m-2}}
				\, \Hat G\du{\r\s\a_1\cdots\a_{2m-2}} I \cr 
= & + \fracm 12 c \left[ \fracm 1{(m-1)!\cdot (2m-3)!! } \right]^2 
					\, \e\du{\m\n}{\r\s} \, F\du{\r\s} I ~~. ~~~~~ ~~~ 
&(\rmA.13) \cr } $$ 
For this to agree with $~F \eqstartext \Tilde F$, we get 
$~c = \left[ \, (m-1) ! \cdot (2m-3)!! \, \right]^2$.  
The remaining components $~{\scst \[\a\b\]}~$ and 
$~{\scst\[\m\a\]}$~ are trivially satisfied.  

The above mechanism for $~D=2m+2$~ is further generalized to 
$~D = 2m+1$~ with the signature $~(-,-,\overbrace{+,+,\cdots,+}^{2m-1})$~ 
with the duality condition 
$$ \li{ & \Hat F\du{\hat\m\hat\n} I 
\eqstar + \fracm1{(2m-1)!} \, 
		\Hat\e\du{\hat\m\hat\n}{\hat\r_1\cdots\hat\r_{2m-1}} 
		\, \Hat G\du{\hat\r_1\cdots\hat\r_{2m-1}} I ~~, ~~~~
		\Hat G\du{\hat\r_1\cdots\hat\r_{2m-1}} I 
		\eqstar + \fracm 12 \, \Hat \e\du{\hat\r_1\cdots\hat\r_{2m-1}} 
			{\hat\m\hat\n}  \, \Hat F\du{\hat\m\hat\n} I ~~. 
			~~~~~ ~~~~~ ~~~ 
&(\rmA.14) \cr } $$ 
The confirmation of $G~$ and $~H\-$BIds is just parallel to the 
$~D=2m+2$~ case.  The ans\" atze for DR is 
$$ \li{ & \Hat G\du{\hat\m_1\cdots\hat\m_{2m-1}} I 
			\eqstar + \fracm{2c'} 3 \Hat F\du{\[ \hat\m_1 \hat\m_2 | } I 
					\Hat P^{(1)}_{| \hat\m_3\hat\m_4 |} \cdots 
					\Hat P^{(m-3)}_{| \hat\m_{2m-5} \, \hat\m_{2m-4} | }
					\Hat Q_{|\hat\m_{2m-3}\hat\m_{2m-2}\hat\m_{2m-1}\]} 					~~, 
&(\rmA.15\rma) \cr 	
& \Hat H\du{\hat\m_1\cdots\hat\m_{2m-2}} I 
			\eqstar + \fracm {2c' g} {2m-1} \Hat F\du{\[ \hat\m_1\hat\m_2| } I 
					\Hat P^{(1)}_{| \hat\m_3 \hat\m_4 | } \cdots 
					\Hat P^{(m-3)}_{| \hat\m_{2m-5} \, \hat\m_{2m-4} | } 
					\Hat Y_{| \hat\m_{2m-3} \hat\m_{2m-2} \]} ~~, 
&(\rmA.15\rmb) \cr 	
& \Hat P^{(k)}_{\hat\m\hat\n} 
			\equiv \Hat\partial_{\hat\m} \Hat X^{(k)}_{\hat\n}
					- \Hat\partial_{\hat\n} \Hat X^{(k)}_{\hat\m}~~, ~~~~ 
\Hat Q_{\hat\m\hat\n\hat\r} \equiv + \Hat\partial_{\hat\m}\Hat Y_{\hat\n\hat\r}
				+ \Hat\partial_{\hat\n} \Hat Y_{\hat\r\hat\m}
				+ \Hat\partial_{\hat\r} \Hat Y_{\hat\m\hat\n}  
				~~,~~~~~ ~~~~~ ~~
&(\rmA.15\rmc) \cr
& \Hat P^{(k)}_{\hat\m\hat\n} = \Hat P^{(k)}_{2k+3, ~2k+4} 
			= - \Hat P^{(k)}_{2k+4, ~2k+3} 	
			= \e^{(k)}_{2k+3,~ 2k+4} 
			= - \e^{(k)}_{2k+4,~ 2k+3} = + 1~~, 
&(\rmA.15\rmd) \cr 
& \Hat Q_{\hat\m\hat\n\hat\r} 
= \Hat Q_{2m-3, 2m-2,2m-1} = \e_{2m-3, 2m-2,2m-1} = + 1 ~~
		 (\hbox{for}~~{\scst \[ \hat\m \hat\n\hat\r \] ~
		 =~\[ 2m-3,2m-2,2m-1\] } ) ~~ , ~~~~~ ~~~~~ ~~  
&(\rmA.15\rme) \cr 
& \Hat F\du{\hat\m\hat\n} I 
= F\du{\m\n} I ~~~~ (\hbox{for} ~~{\scst \hat\m~=~\m,~\hat\n~=~\n}) ~~,  
&(\rmA.15\rmf) \cr 
& \Hat \e\,{}^{\hat\m_1\cdots\hat\m_{2m+1}}  
= \e^{\m\n\r\s} \, \e^{\a_1\cdots\a_{2m-3}} 
		= \e^{\m\n\r\s} \, \e_{(1)}^{\[ \a_1\a_2 | } \cdots 
		\, \e_{(m-3)}^{| \a_{2m-7}\a_{2m-6} | } 
		\e^{|\a_{2m-5} \a_{2m-4} \a_{2m-3} \]} ~~.  
& (\rmA.15\rmg) \cr } $$  
The totally antisymmetric constant tensor $~\e^{\a\b\g} $~ is 
for the last three coordinates in $~D=2m+1$.  				
The satisfaction of the duality (A.14) fixes the constant  
~$c' = \[ (m-3)! \cdot (2m-7) !! \]^2$.

\newpage


\def\texttts#1{\small\texttt{#1}}

\immediate\closeout\rfile\writestoppt
\baselineskip=12.5pt\centerline{{\bf References}}
\font\smallerfonts=cmr10 \font\it=cmti10 \font\bf=cmbx10%
\bigskip{\smallerfonts{%
\parindent=18pt\escapechar=` \input refs.tmp\vfill\eject}}


\vfill\eject

\end{document}